\documentclass[namedreferences]{SolarPhysics}

\usepackage[hyperref,optionalrh]{spr-sola-addons} 
\usepackage{graphicx}        
\usepackage{breakurl}
\usepackage{courier}         
\usepackage[usenames,dvipsnames]{color}             
\usepackage{url}             

\newcommand{\p}[1]{{\textbf{#1}}}

\newcommand{\pder}[2]{ \frac{\partial #1}{\partial #2} }
\newcommand{\sign}{\mathop{\mathrm{sign}}}

\newcommand{\rmd}{ {\mathrm d} }
\renewcommand{\vec}[1]{\mathbfit{#1}}

\newcommand{\arcsec}{$''$} 
\newcommand{\degree}{^\circ} 

\newcommand{\BE}{\begin{equation}}
\newcommand{\EE}{\end{equation}}
\newcommand{\BA}{\begin{eqnarray}}
\newcommand{\EA}{\end{eqnarray}}
 \newcommand{\fig}[1]{Figure~\ref{fig:#1}}

 \newcommand{\figss}[2]{Figures~\ref{fig:#1}\,--\,\ref{fig:#2}}
 \newcommand{\sect}[1]{Section~\ref{sec:#1}}
 \newcommand{\append}[1]{Appendix~\ref{app:#1}}
 \newcommand{\sects}[2]{Sections~\ref{sec:#1} and \ref{sec:#2}}
 
 \newcommand{\eq}[1]{Equation~(\ref{eq:#1})}
 \newcommand{\eqs}[2]{Equations~(\ref{eq:#1}) and (\ref{eq:#2})}

\newcommand{\eg}{\textit{e.g.}}
\newcommand{\etal}{\textit{et al.}}
\newcommand{\ie}{\textit{i.e.}}

\newcommand{\vs}{\textit{vs.}}

\newcommand{\asy}{{\rm asy}}
\newcommand{\Bt}{B_{\rm \theta}}
\newcommand{\Bp}{B_{\rm \phi}}  
\newcommand{\Brho}{B_{\rm \rho}}
\newcommand{\DelAbsphic}{\Delta|\phi_{\rm c}|}
\newcommand{\DelAbstauc}{\Delta|\tau_{\rm c}|}
\newcommand{\Delphic}{\Delta \phi_{\rm c}}
\newcommand{\Deltauc}{\Delta \tau_{\rm c}}
\newcommand{\Fmax}{F_{\rm max}}

\newcommand{\Nt}{N_{\rm t}}
\newcommand{\Nto}{N_{\rm t,0}}

\newcommand{\phic}{\phi_{\rm c}}

\newcommand{\tauc}{\tau_{\rm c}}

\newcommand{\taucFL}{\tau_{\rm c, L. fit}}

\newcommand{\taucMean}{\overline{\tau_{\rm c}}}
\newcommand{\taucMax}{\tau_{\rm c,max}}
\newcommand{\te}{t_{\rm end}}

\newcommand{\ti}{t_{\rm init}}
\newcommand{\tm}{t_{\rm Fmax}}
\newcommand{\torus}{torus}
\newcommand{\ts}{t_{\rm start}}
\newcommand{\up}{\vec{\widehat{u}}_{\rm \phi}}
\newcommand{\uR}{\vec{\widehat{u}}_{\rm R}}
\newcommand{\urho}{\vec{\widehat{u}}_{\rm \rho}}
\newcommand{\ut}{\vec{\widehat{u}}_{\rm \theta}}

\newcommand{\uz}{\vec{\widehat{u}}_{\rm z}}

\begin{document}

\begin{article}

\begin{opening}

\title{Properties of Magnetic Tongues over a Solar Cycle}

\author[addressref={1},corref,email={mpoisson@iafe.uba.ar}]{\inits{M.}\fnm{Mariano }\lnm{Poisson}}
\author[addressref={2},corref,email={}]{\inits{P.}\fnm{Pascal }\lnm{D\'emoulin}}
\author[addressref={1},corref,email={}]{\inits{M.}\fnm{Marcelo }\lnm{L\'opez Fuentes}}
\author[addressref={1,3},corref,email={}]{\inits{C. H. }\fnm{Cristina~H.~}\lnm{Mandrini}}

\runningauthor{M. Poisson \etal }
\runningtitle{Properties of Magnetic Tongues over a Solar Cycle}

\address[id={1}]{Instituto de Astronom\'\i a y F\'\i sica del Espacio (IAFE), CONICET-UBA, Buenos Aires, Argentina}

\address[id={2}]{Observatoire de Paris, LESIA, UMR 8109 (CNRS), F-92195 Meudon Principal Cedex, France}

\address[id={3}]{Facultad de Ciencias Exactas y Naturales (FCEN), UBA, Buenos Aires, Argentina}

\begin{abstract}
The photospheric spatial distribution of the main
magnetic polarities of bipolar active regions (ARs) presents during
their emergence deformations are
known as magnetic tongues. They are attributed to the
presence of twist in the toroidal magnetic  flux-tubes that form the
ARs. The aim
of this article is to study the twist of newly emerged ARs from the evolution
of magnetic tongues observed in photospheric line-of-sight
magnetograms. We apply the
procedure described by Poisson et al. (2015, \textit{Solar Phys.} \textbf{290}, 727) to ARs
observed over the full Solar Cycle 23 and the beginning of Cycle 24.
Our results show that the hemispherical
rule obtained using the tongues as a proxy of the twist has a weak
sign-dominance
(53\,\% in the southern hemisphere and 58\,\% in the northern hemisphere). By defining
the variation of the tongue angle, we characterize the strength of the magnetic
tongues during different phases of the AR emergence. We find that there is
a tendency of the tongues to be stronger during the beginning of the
emergence and to become weaker as the AR reaches its maximum magnetic  flux.
We compare this evolution with the emergence of a toroidal  flux-rope model
with non-uniform twist. The variety of evolution of the tongues in
the analyzed ARs
can only be reproduced when using a broad range of twist profiles, in
particular having
a large variety of twist gradient in the direction vertical to the photosphere.
Although the analytical model used is a special case, selected to minimize the complexity of the problem, the results obtained set new observational constraints to theoretical models of flux-rope emergence that form bipolar ARs.
\end{abstract}


\end{opening}


\section{Introduction}
 \label{sec:Introduction} 

The study of the solar-cycle properties is the basis for understanding how the solar dynamo produces and amplifies the magnetic field in the solar interior.  In the last five decades, several models have proposed a dynamo mechanism located at the bottom of the convective zone (CZ). In this scenario, the magnetic flux at the base of the CZ is amplified and distorted by differential rotation and convection and, finally, it  is destabilized by a buoyant-instability process. Magnetohydrodynamic (MHD) numerical simulations show the way that this instability creates coherent magnetic-tubes that rise from the deep layers of the CZ and manifest as the emergence of solar active regions (ARs) observed at photospheric heights \citep{Fan09r}. An important prediction of these simulations is that the emerging structures should form twisted flux tubes or flux ropes (FR) in order to maintain their consistency during the transit through the turbulent CZ.

In this view, ARs are the consequence of the emergence of FRs. Other mechanisms, however, have been proposed to explain the formation of ARs (see \citet{Cheung14} and references therein). 
There is much observational evidence of twist in ARs  \citep[\ie\ sunspot whorls, non-potentia\-lity of coronal loops, prominences, X-ray sigmoids; see the review of ][]{Pevtsov14}. ARs with complex magnetic-field distribution, associated with highly twisted FRs, are more productive in terms of flares and coronal mass ejections (CMEs) due to the amount of free magnetic energy and helicity stored in their structures \citep{Kusano04,Liu08,Tziotziou12,Szajko13}. 

The tendency of magnetic helicity to have a different sign in each solar hemispheres was first proposed by Hale from the evidence of vortical patterns in the chromospheric fibrils surrounding the ARs sunspots \citep{Hale25}. According to this ``hemispherical rule'', there is a predominancy of positive (negative) helicity in the southern (northern) hemisphere. The strength of the rule has been tested using several estimations of the magnetic and current helicity \citep{Bao00,LaBonte07,Pevtsov08,Liu14}. However, a wide range of variation is observed in the rule, \eg\, it is weaker for ARs ($\approx 60$\,\% -- 70\,\%) than for quiescent filaments \citep[$\approx 80$\,\%;][]{Wang13,Pevtsov14}. The significant scatter exhibited by the hemispheric rule implies that turbulence in the convection zone may play an important role in the generation of the observed chirality trends \citep{Longcope98,Nandy06}. 

\citet{Lopez-Fuentes00} reported a proxy of the twist associated with the deformation of the magnetic polarities observed in line-of-sight magnetograms. These observed features, called magnetic tongues (or tails), are produced by the azimuthal field component of the emerging FR projected on the line-of-sight. The magnetic-flux distribution due to the magnetic tongues is directly related to the sign of the twist in an emerging AR, so the deformation or elongation of the magnetic polarities can be used as a proxy for the sign of the helicity. \citet{Luoni11} computed the elongation of the AR polarities and the evolution of the polarity inversion line (PIL) to characterize the strength of the magnetic tongues.  They found that the twist sign inferred from the observed magnetic tongues is consistent with the helicity sign deduced from other proxies (photospheric-helicity flux, sheared coronal loops, sigmoids, flare ribbons, and/or the associated magnetic cloud). 

\citet{Poisson15} introduced a systematic method to quantify the effect of the magnetic tongues by studying the PIL evolution during the emergence of ARs. This less user-dependent procedure consists in the systematic computation of a linear approximation of the PIL in between the strong magnetic polarities. This is done by minimizing the opposite-sign magnetic-flux component on each side of the computed PIL (see \sect{Characterizing}). From the acute angle between the computed PIL and the line orthogonal to the AR bipole axis [$\tau$] we estimated the average number of turns in the sub-photospheric emerging flux rope, assuming that it can be represented by a uniformly twisted half torus.  We found that the number of turns [$\Nt$] is typically below unity; then, sub-photospheric flux-ropes have in general a low amount of twist.

In a more recent article \citet{Poisson16} compared $\Nt$ with the twist of simple bipolar ARs calculated from linear force-free field extrapolations of their line-of-sight (LOS) magnetic field to the corona.  The signs of the twist obtained with both methods are consistent. 
Moreover, we found a linear relation between $\Nt$ computed at the photospheric level from the tongues and the number of turns obtained from coronal field modeling. 

In this article, we use the procedure described by \citet{Poisson15} to explore the properties of the twist of emerging ARs during a complete solar cycle. Our main aim is to search for any dependence of the parameters (\ie\ $\Nt$, $\tau$) characterizing magnetic tongues on the different phases of the cycle and to expand the previous results to a larger statistical sample. Moreover, we quantify the strength of the tongues in the different stages of the AR emergence. This provides further information about how the twist is distributed in the FR.  In \sect{Observations}, we describe the data used and the selected sample of ARs. We also define the procedure to derive the tongue characteristics.  In \sect{Properties}, we study the properties of the magnetic tongues; in particular, along the solar cycle. In \sect{FRmodel}, we compare the evolution of the PIL angle during an AR emergence with the emergence of a FR model that we develop to interpret the variety in the evolution of the magnetic tongues. In particular, we show that a broad range of twist profiles is required to interpret the observations.  Finally, in \sect{Conclusions}, we summarize our results and conclude.

\section{Observations and Methods}
 \label{sec:Observations}

\subsection{Active Regions Studied}
 \label{sec:StudiedAR}
 
We used line-of-sight magnetograms from the {\it Michelson Doppler Imager} \citep[MDI:][]{Scherrer95}, onboard the {\it Solar and Heliospheric Observatory} (SOHO). The full-disk magnetograms consist of 1024 $\times$ 1024 pixel arrays and are calibrated at level 1.8. \p{These 96-minute cadence maps are constructed using either one-minute or  five-minute averaged magnetograms. The data series consist of 15 magnetograms {\it per} day with a spatial resolution of 1.98\arcsec and an error in the flux density of 16 G or  9 G {\it per} pixel, respectively.} 


We selected ARs that appeared on the solar disk during the full operational period of the MDI instrument, from the beginning of Solar Cycle 23 ($\approx$ July 1996) until the beginning of Cycle 24 (around January 2010). We systematically searched for ARs with dominantly bipolar magnetic-field configurations and low background flux in this period of time ($\approx$ 13 years). These are the same criteria used by \citet{Poisson15}. The selection was made by examining SOHO/MDI images provided by the Helio-viewer website (\url{www.helioviewer.org}). After selecting the ARs in this way, we retrieved the corresponding original data from the MDI instrument data base.

Since we are interested in the emergence phase of the ARs, we need to track their evolution through the largest possible part of their transit, but minimizing the projection and foreshortening effects. Therefore, we selected ARs that started emerging between 30$^{\circ}$ East and the central meridian (CM). We found 187 well-isolated bipolar ARs emerging in areas devoid of significant background field during this period of time. The constraints on the characteristics of the selected ARs limit the statistical size of our sample; however, our aim is to have a set of clearly observed emerging ARs. 

We processed the full-disk magnetograms using standard Solar-Software tools. We first transformed the magnetic-flux density measured in the observer's direction to the solar radial direction neglecting the contribution of the components on the photospheric plane. This is a small correction as we selected ARs that are close to the disk center (the effects of this transformation were studied, \eg, by \citet{Green03}). Then, we proceeded to rotate the magnetograms to the date when the AR was located at the CM; this rotation reduces the foreshortening effect which is present away from the CMP (a small effect since we analyze ARs closer than 30$\degree$ from
CMP). We removed from the set any magnetogram with evidence of bad quality measurements. Finally, we selected a large sub-region of the magnetograms where the AR is located during the emergence phase and later evolution. The selected sub-region is the same for all of the magnetograms containing the observed AR. All of the processed data can be easily stored in a three-dimensional array (two spatial dimensions for each map and one for the temporal evolution).

Next, we chose rectangular boxes of variable size encompassing the AR polarities during all of the emergence phase. Movies for each AR were made to verify that the variable-size box included all of the magnetic flux of the AR at all times. All of the AR parameters were computed considering only the pixels inside these rectangular regions from the AR's first emergence until the maximum flux was reached and/or projection and foreshortening effects were too important (when the distance to the CM passage was larger than $30\degree$).

\subsection{Characterizing Magnetic Tongues}
 \label{sec:Characterizing}

\begin{figure}[t]
\begin{center}
\includegraphics[width=\textwidth]{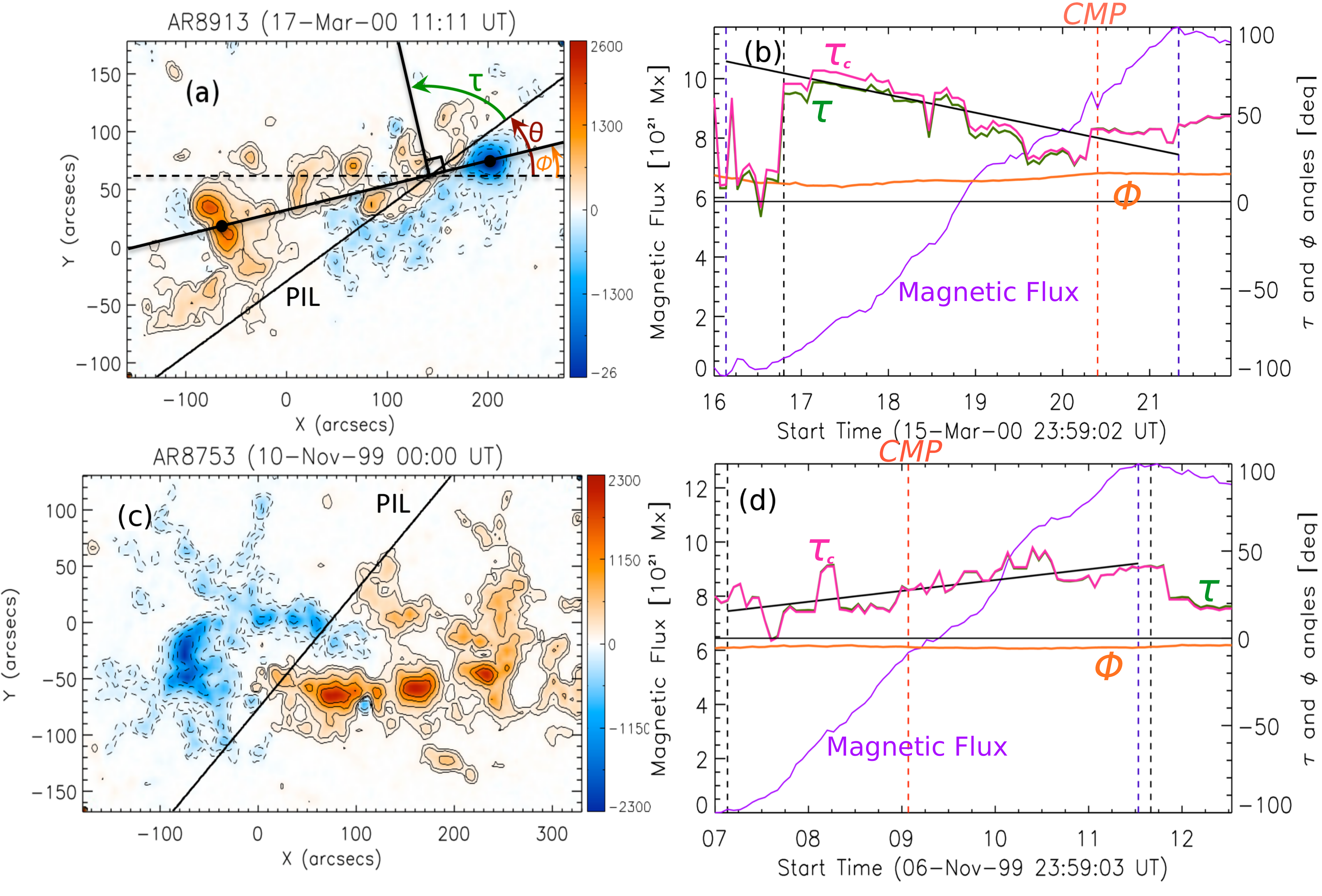}
\caption{(a,c) Two magnetogram examples of emerging ARs observed by MDI (see Electronic Supplementary Materials \href{run:./movie1a.mp4}{\sf{Movie1a}} and \href{run:./movie1c.mp4}{\sf{Movie1c}}). The continuous (dashed) lines are isocontours of the positive (negative) magnetic field component normal to the photosphere, drawn for 100, 500, 1000, and 1500 G (and the corresponding negative values). The red- and blue-shaded regions indicate positive and negative polarities, respectively.  The thin-black continuous line is the best straight PIL computed by the method described in \sect{Characterizing}.
In (a) the mean positions of the polarities are marked by black dots and a straight line is added to define the bipole direction. Its orthogonal direction is also drawn. The horizontal-dashed line is along the East--West direction. The $\theta$, $\phi$, and $\tau$ angles are defined using these lines.
   (b,d) Evolution of the magnetic flux (violet-continuous line), the tongue angle ($\tau$, green line), the corrected tongue angle ($\tauc$, pink line), and the tilt angle ($\phi$,  orange line) for the ARs in panels a and c, respectively.   A linear fit to the $\tauc$ curve is drawn with a continuous-black line. Pairs of vertical-dashed black lines mark the time periods of emergence and pairs of vertical-dashed purple lines those where we perform the least-square's fit of $\tauc$ (in (b) the end times and in (d) the beginning times are superposed).  The orange-vertical-dashed line indicates the AR CM passage. In panel d the $\tau$-curve is mostly superposed by the one of $\tauc$ (negligible correction from \eq{tauc}).
   }
 \label{fig:example}
\end{center} 
\end{figure}

As we described in \sect{Introduction}, magnetic tongues are evident during the emergence of ARs (at least the ones that are not too complex and where multipolarities can mask or distort new emergences). For twisted FRs, the projection of the magnetic-field components on the observer's direction results into an elongation of the AR polarities that can be seen in MDI magnetograms \citep{Luoni11}. We used the polarity inversion line (PIL) method described by \citet{Poisson15} to characterize the deformation of the polarities along the AR evolution. Although the PIL can be a complicated curved line, the method finds the coefficients $a,b,c$ to approximate the PIL to a straight line by minimizing the function $D(a,b,c)$ defined as:
  \BA
   \label{eq:def-dif}
   D(a,b,c) = \int_{x,y} |\sign (a x + b y + c)~|B_o| - B_o|^2\; \rmd x \; \rmd y,
  \EA 
\noindent where $B_o = B_o(x,y)$ is the observed field at location $(x,y)$ and the integration is done over a rectangular region surrounding the AR (avoiding the flux not belonging to the AR).  The minimization of $D(a,b,c)$ finds the straight line that separates in the best possible way the AR in two distinct polarity regions.  The main AR field is not contributing to $D(a,b,c)$ as its contribution cancels in the integrand by construction. Only the polarities in the vicinity of the PIL have a variable contribution when the coefficients $a,b,c$ are changed around the values minimizing $D(a,b,c)$.  A further description of the method has been given by \citet{Poisson15}.  Finally, we define the PIL angle [$\theta$] as the angle between the straight PIL and the East--West direction (see \fig{example}a).
 
We also compute the tilt angle of the AR. To do so, we use the line that joins the flux-weighted centers (which we call barycenters) of the positive and negative magnetic polarities \citep{Lopez-Fuentes00} and we give the name of the bipole vector to the oriented axis that joins the following to the leading barycenters. The tilt angle [$\phi$] of an AR is defined as the angle between the East--West direction and the bipole vector.

Next, we define the tongue angle [$\tau$] as the acute angle between the PIL direction and the direction orthogonal to the AR bipole vector as: 
  \BE  \label{eq:tau}
   \tau = \phi - \theta + 90\degree  \,.
  \EE
Since $\phi$ lies between $-90\degree$ and $90\degree$ and the PIL direction is defined modulo $180\degree$, we select  $\theta$ so that $\theta-\phi$ lies between $0\degree$ and $180\degree$. This implies that $\tau$ lies between $-90\degree$ and $90\degree$, and $\tau$ has the same sign as the twist that is inferred from the shape of the tongues \citep{Poisson15}.

Since for some ARs a large amount of magnetic flux  is present in the magnetic tongues, their temporal evolution has an effect on the bipole tilt. That is why we define a corrected tilt angle considering that the flux-tube axis direction can be inferred from the tilt angle when the tongues have retracted and before another emergence or dispersion further modifies the bipole tilt. We consider the time of maximum flux as the one satisfying both conditions, as inferred from the analysis of the evolution of ARs (see examples in Figure 1 and Figures 4 to 10 of
\citet{Poisson15}). Then, we define the corrected angles [$\tauc$,$\phic$] as
  \BA 
  \tauc (t) &=& \tau (t) - \phic (t)  \label{eq:tauc}\, , \\
  \phic (t) &=& \phi (t) - \phi_{\rm max~flux} \label{eq:phic}\, ,
  \EA
$\tauc (t)$ is the tongue angle with respect to the direction defined by $\phi_{\rm max~flux}$ (rather than $\phi$) and $\phic(t)$ is the variation of the tilt angle.
In a following step, we compute an average of $\tauc (t)$ [$\taucMean$] and a maximum value [$\taucMax$] within the emerging time period as defined in the next section.

\begin{figure}[t]
\begin{center}
\includegraphics[width=\textwidth]{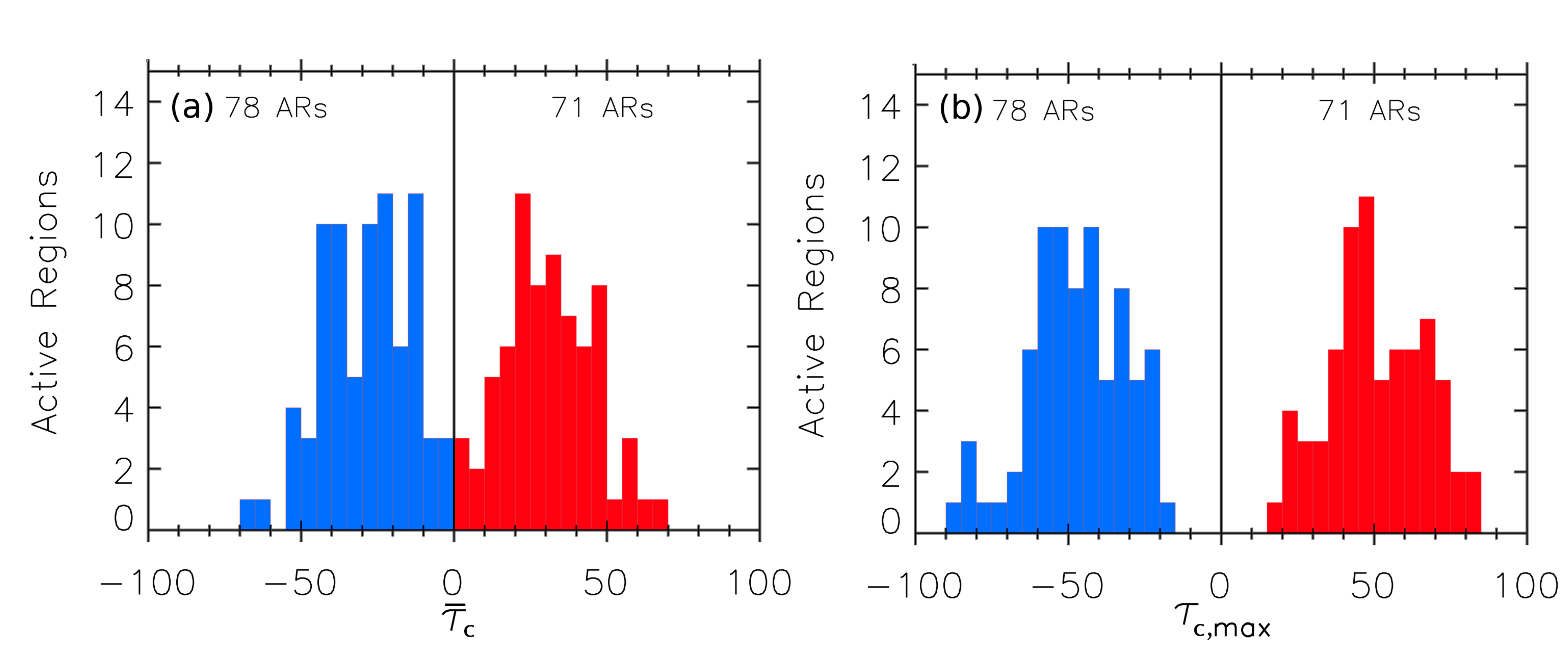}
\caption{Histogram of the mean tongue angle [$\taucMean$] and its signed maximum [$\taucMax$] for 149 ARs; these are the ones for which the computation of these parameters could be done during more than $50~\%$ of the AR emergence time (ARs represented with filled triangles 
in \fig{AR(latitude,time)}).
}
 \label{fig:histo-tau}
\end{center} 
\end{figure}

In \fig{example} we present the results derived for two observed ARs with positive twist. We plot the AR tilt [$\phi$] (orange line), the magnetic flux (violet continuous line), $\tau$ (green line), and $\tauc$ (pink line) during the emergence time. The plots show that there is a characteristic evolution of these AR parameters associated with the formation of the tongues. In particular, the sign of $\tauc$ is well defined during all of the emergence, as shown in \fig{example}b,d, and it corresponds to positively twisted flux tubes \citep[see, \eg\,, Figure 1a of][]{Poisson15}. In the movies for both ARs, included as Electronic Supplementary Material, we can see a continuous evolution of the PIL direction and the tilt as the magnetic tongues retract. For AR 8913 (\fig{example}b), the tilt variation is low during the emergence and it quickly becomes stable once the maximum flux is reached. The effect of the tongues on the tilt inclination can be seen in the difference between $\tau$ and $\tauc$ in \fig{example}b.  For AR 8753 the magnetic flux in the tongues is small, this implies $\tauc \approx \tau$ (\fig{example}d).   Finally, $|\tauc|$ is significant for both ARs and takes typical values (20\,--\,50 $\degree$, see \fig{histo-tau}a).

\subsection{Analyzing the Temporal Evolution of the Tongue Angle }
 \label{sec:Analyzing}

To study the evolution of the tongue angle, we define the emergence-time range as $[\ts , \tm ]$ between the beginning of the emergence and the maximum value of the magnetic flux. The time interval during which the measurements of $\tauc$ are done is $[\ti , \te ]$ (see the black-dashed lines in \fig{example}b,d). Due to data gaps or too strong fluctuations because of parasitic bipole emergences (mostly at the beginning of the emergence time) or to a significant secondary emergence, this last time interval could be shorter than the full AR emergence duration: 
  \BE \label{eq:times}
  \ts \leq \ti < \te \leq \tm \, .
  \EE

We characterize the evolution of the magnetic tongues during the emergence phase using a linear fit of $\tauc$ within the interval $[\ti , \te]$ as given by 
  \BE  \label{eq:linearFit}
  \taucFL (t)= a+b ~t,
  \EE
where $a$ and $b$ are the $y$-intercept and the slope in the linear fit.  Examples of these linear fits are shown in \fig{example}b,d. These fits capture the mean evolution of $\tauc$, filtering the fluctuations of the emergence. However, in 38 over 187 ARs the fluctuations were dominant so that a linear fit had no meaning and was not performed. In the rest of the ARs a linear trend is dominant in more than 50 \% of the emerging time. We have also tried a second-order-polynomial fit but obtained no further significant information.
 All of the values derived from the $\tauc$ evolution, such as the average and the maximum  of $\tauc$ [$\taucMean$, $\taucMax$], are also computed within this time interval.   


 We can repeat the same procedure for the fit of $\tauc$ within the time interval $[\ti , \te ]$, but redefining $\tauc$ as a function of the normalized magnetic flux [$F/\Fmax$, where $\Fmax$ is the maximum flux]. We perform a linear fit of $\tauc$: 
  \BE  \label{eq:linearFitF}
  \taucFL (F) = a'+b' ~F/\Fmax \, .
  \EE
Then, we define the signed change of $|\tauc |$ as 
  \BE \label{eq:dtau}
  \DelAbstauc = \sign(\taucMean)~b' \, .
  \EE
In this way we remove the dependence of $\DelAbstauc$ on the rate and the duration of the emergence, obtaining a more comparable set of parameters for all the ARs. The amount of background flux (or any disturbing flux) is very case-dependent, limiting the time interval $[\ti ,\te]$ to a variable fraction of $[\ts , \tm]$. Because $F/\Fmax$ varies in the interval $[0,1]$ in the definition of $\DelAbstauc$, given by \eq{dtau}, we continue computing $\DelAbstauc$ within $[\ts , \tm ]$  
to minimize the effect of various perturbations to the flux. This assumes that $\tauc (t)$ follows the same trend during all the AR emergence phase. 

We include $\sign(\taucMean)$ in the $\DelAbstauc$ definition, given by \eq{dtau}, to compare directly the ARs with positive and negative $\tauc$. Then, for both cases, when $\DelAbstauc > 0$ the tongues are weaker at the beginning of the emergence but become stronger as the AR reaches the maximum flux and {\it vice versa}. \fig{example} shows two examples with the same sign of $\taucMean$ and different signs of $\DelAbstauc$ (for AR 8913 $\DelAbstauc = -54\degree$ and for AR 8753 $\DelAbstauc = 28\degree$).

\section{Properties of Magnetic Tongues}
 \label{sec:Properties} 

\subsection{How Important is the Effect of Magnetic Tongues?}
 \label{sec:Strong}

Magnetic tongues are created by the azimuthal magnetic-field component of the FR projected on the local vertical ($B_{z}$, and more generally along the line of sight).   However, both azimuthal and axial components in a magnetic FR contribute to $B_{z}$ in different proportions, which vary continuously across the AR magnetogram.
This implies that the contribution of the azimuthal flux cannot be estimated. We can only estimate the total axial flux when the FR cross-section has fully emerged through the photosphere and the magnetic tongues have retracted.

Still, the PIL location results from a balance between the azimuthal and axial magnetic-field components and its orientation characterizes the importance of the tongues.  Tongues are absent for $\tauc \approx 0$ and become stronger as $|\tauc|$ is closer to $90 \degree$, a limit where the azimuthal flux totally dominates the axial flux, \ie\ when the FR is very twisted.  A coherent evolution of $\tauc$ is present in the majority of ARs (two examples are shown in \fig{example}). However, in some ARs, 38 of 187, $\tauc (t)$ has strong variations because extra background or emerging flux is present. We analyze below only the 149 ARs where a linear fit of $\tauc$ is a good approximation to its evolution in more than 50\% of the AR-emergence duration.

Magnetic tongues are clearly observed in the large majority of emerging ARs that follow the selection criteria described in \sect{StudiedAR}, as shown from the significant values of $\taucMean$ (\fig{histo-tau}a) and, even more, if we analyze the maximum value of $\tauc$ (\fig{histo-tau}b).  The mean of $|\taucMean| \approx 30\degree$ and the one of $|\taucMax| \approx 50\degree$ confirms the results of \citet{Poisson15} for a larger sample of ARs: an isolated bipolar AR is typically formed by an emerging FR with a significant twist.

The sign of $\tauc$ is a direct proxy for the sign of the magnetic helicity \citep{Luoni11}. The distributions of both $\taucMean$ and $\taucMax$ (\fig{histo-tau}) are very similar within statistical fluctuations for positive and negative $\tauc$. The mean of $\taucMean$ is approximately $31\degree$ for the positive distribution and $-29\degree$ for the negative one, both having the same standard deviation ($\approx 14\degree$). We find the same correspondence for $\taucMax$ distributions. This implies that emerging active regions (following the criteria described in \sect{StudiedAR}) with positive and negative twist have similar statistical properties for $\taucMean$ and $\taucMax$.

\begin{figure}[t]
\begin{center}
\includegraphics[width=0.9\textwidth]{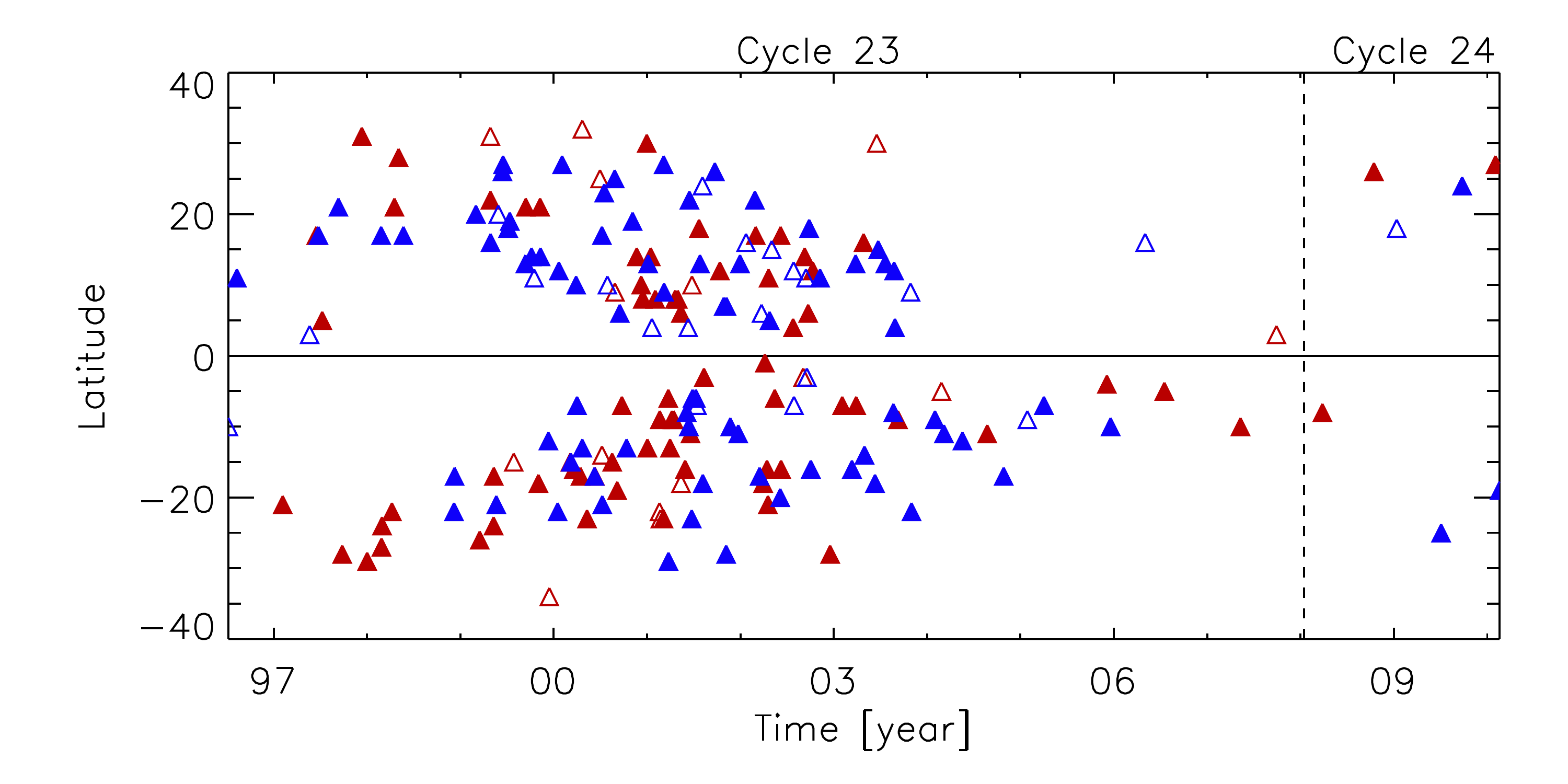}
\caption{AR latitudes for the full Solar Cycle 23.  Positive (negative) helicity sign, as deduced from magnetic tongues, is indicated with a red (blue) triangle. Filled triangles mark ARs where the period of time in which we perform the computations is more than $50~\%$ of the total emergence (149 ARs for a total of 187 ARs).  Solar cycle 23 is fully covered.
}
 \label{fig:AR(latitude,time)}
\end{center} 
\end{figure}

\begin{figure}[t]
\begin{center}
\includegraphics[width=\textwidth]{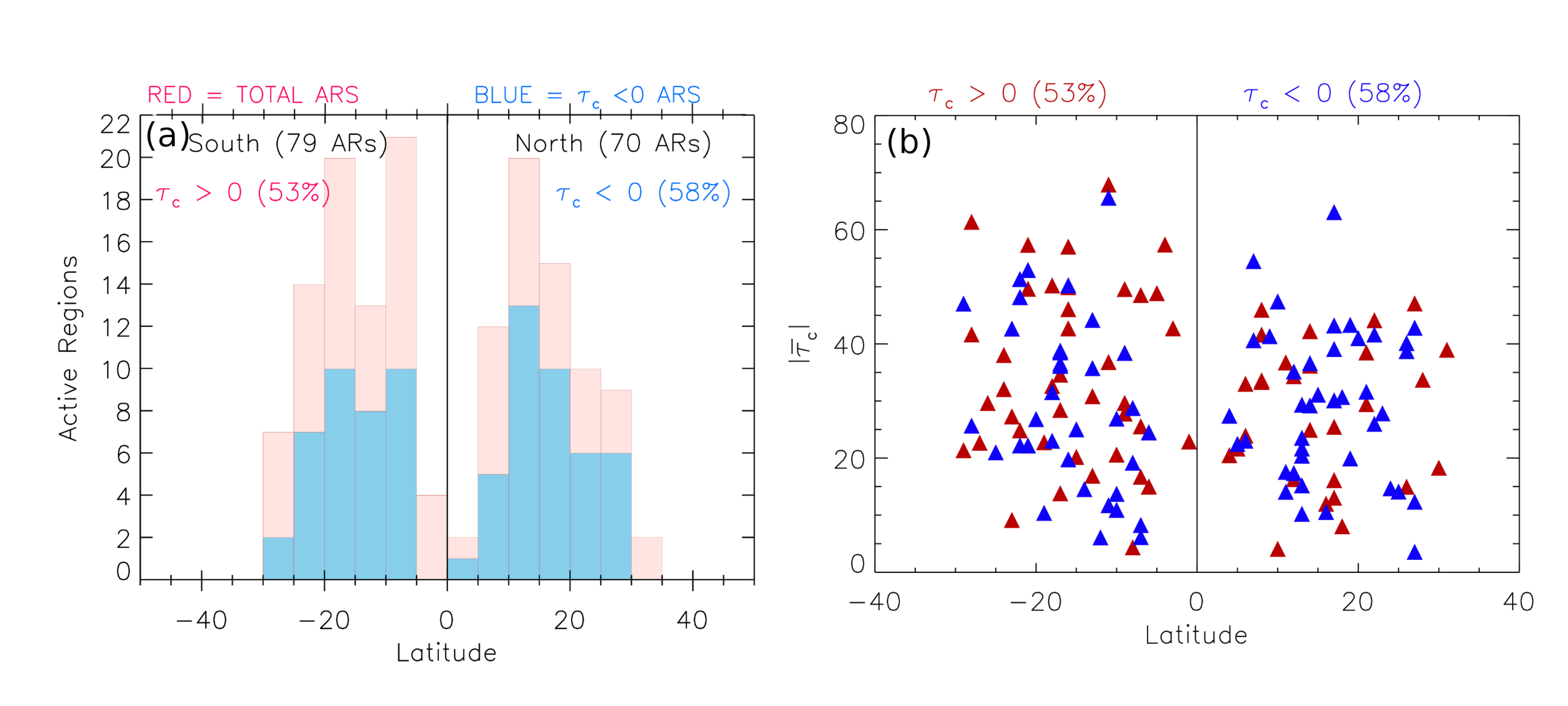}
\caption{Helicity-sign dependence in function of latitude.
   (a) The pink histogram, in the background, shows the latitude for all the studied ARs. The blue histogram shows negative twisted ARs ($\taucMean <0$), then the visible pink part of the total histogram shows the positive twisted ARs.
   (b) Mean absolute value of the tongue angle, $|\taucMean |$, \vs\ latitude for positive (negative) twist ARs shown in red (blue). $53$\,\% of ARs from the southern hemisphere have positive $\taucMean$ and $58$\,\% from the northern hemisphere have negative $\taucMean$.
}
 \label{fig:helicity-sign}
\end{center} 
\end{figure}

\begin{figure}[t]	
\begin{center}
\includegraphics[width=0.8\textwidth]{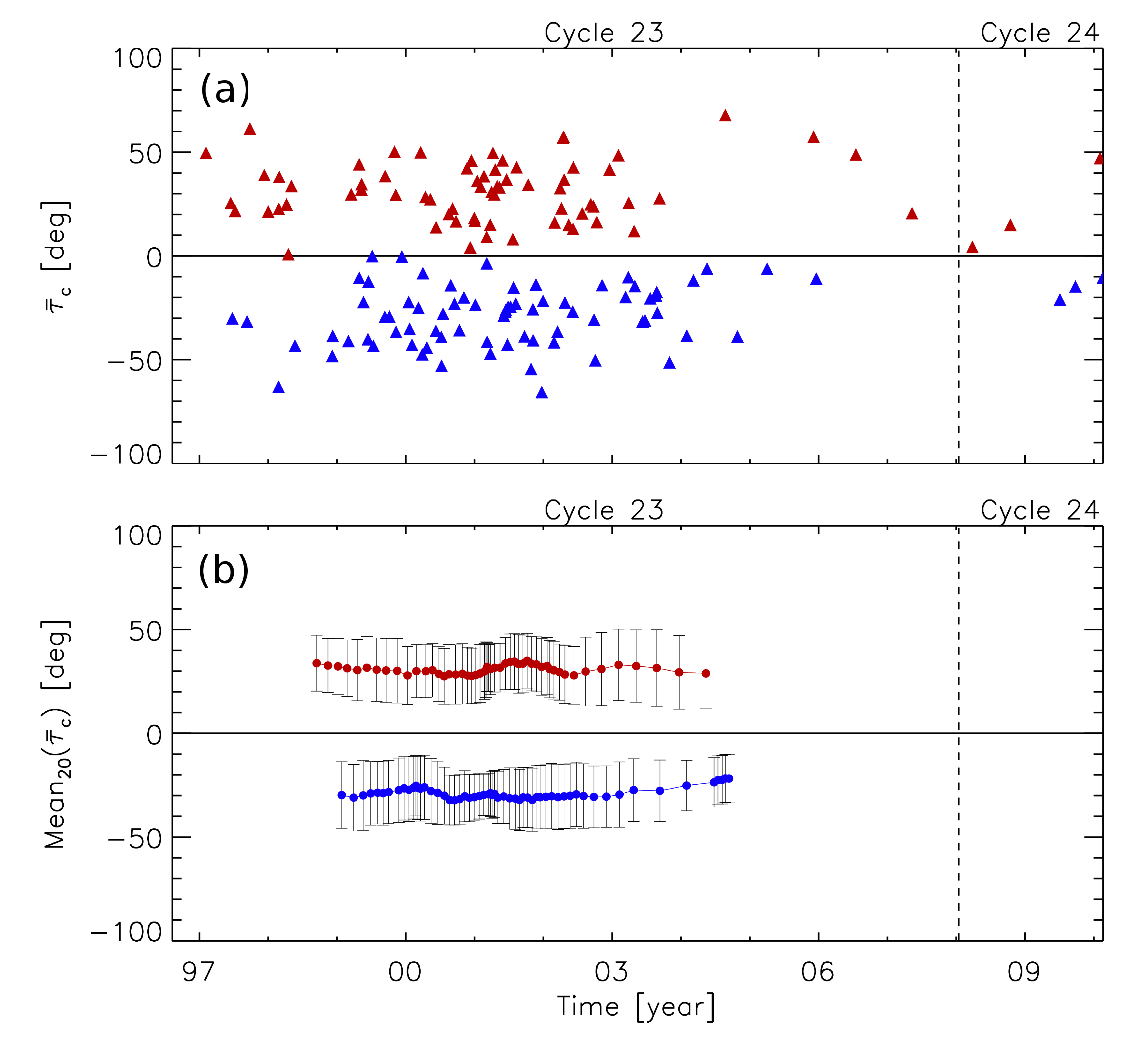}
\caption{
  (a) Mean tongue angle [$\taucMean$] shown for the 149 ARs observed mostly during Solar Cycle 23 (red: $\taucMean>0$, blue $\taucMean<0$). 
  (b) 20-point running mean for $\taucMean$ along the cycle. Error bars represent the standard deviation of each 20-point sample. 
}
 \label{fig:AR(tau,time)}
\end{center}
\end{figure}

\subsection{Distribution of the Tongues with Latitude and during the Solar Cycle}
 \label{sec:Latitude-Cycle}

The distribution of the studied ARs in a latitude {\vs} time plot shows the classical butterfly diagram (\fig{AR(latitude,time)}).  Indeed, our selection of ARs is based on their distance to the CM passage and the absence of significant magnetic-field background at the AR emergence; therefore, the selection should not be biased in latitude and time apart from a small number of ARs selected around solar maximum when a stronger nesting effect is present. During the solar maximum more emerging ARs were rejected because they appear either too close to, or even within, a pre-existing AR or they are too complex; so finally they are not satisfying the selection criteria described in \sect{StudiedAR}. Nevertheless, the number of selected ARs is still larger {\it per} year around the solar maximum.    

The sign of $\taucMean$ is mixed both in latitude and time with no clear dominant helicity sign in each hemisphere (\fig{AR(latitude,time)}).  \fig{helicity-sign}a confirms quantitatively that the helicity-sign hemispherical-rule is weak.  This result, derived for isolated bipolar ARs, agrees with those of other studies (see \sect{Introduction}) in which the segregation by hemisphere of the magnetic helicity-sign is weak for young magnetic structures, getting stronger as the structures are older.  For example, the hemispherical rule ($H<0$ in the northern hemisphere and $H>0$ in the southern one) typically has an imbalance in the range 60\,--\,70\,\% for the magnetic field of mature ARs, with an imbalance higher than 90\,\% for quiescent filaments outside ARs \citep{Pevtsov02b,Pevtsov08}. 
 
The previous results are not due to weak $|\tauc|$ values, which could be more sensitive to noise, because the mixture of signs is equally present for all $|\taucMean|$ strengths, as well as for all latitudes (\fig{helicity-sign}b).  There is also no evidence of the evolution of the value of $|\tauc|$ during the solar cycle as the mean value and the dispersion of $|\tauc|$ is mostly independent of time (\fig{AR(tau,time)}).  More precisely, the temporal variation of $|\tauc|$ is only of the order of $10\degree$ and within the standard error bars (\fig{AR(tau,time)}b). 

The above results correspond to the subset of 149 ARs for which at least 50\,\% of the observed emergence can be linearly fitted (see \sect{Analyzing}). Similar results are obtained for the full set of 187 ARs for the helicity-sign hemispherical rule ($\approx$54\,\% in the southern hemisphere and $\approx$60\,\% in the northern hemisphere), showing that the restriction of the 50\,\% threshold, which only affects 20\,\% of the AR sample, has no impact on the weakness of the helicity-sign dependence. Increasing the threshold of the linear fit to 75\,\% of the AR emergence, reduces the sample by 
more than 45\,\% while the helicity-sign dependence is sligthly modified (it turns out to be $\approx$ 49\,\% in the southern hemisphere and $\approx$60\,\% in the northern hemisphere). Therefore, the 50\,\% threshold chosen for the computation does not significantly affect the results, and it still guarantees a large statistical sample.

\begin{figure}[ht]
\begin{center}
\includegraphics[width=\textwidth]{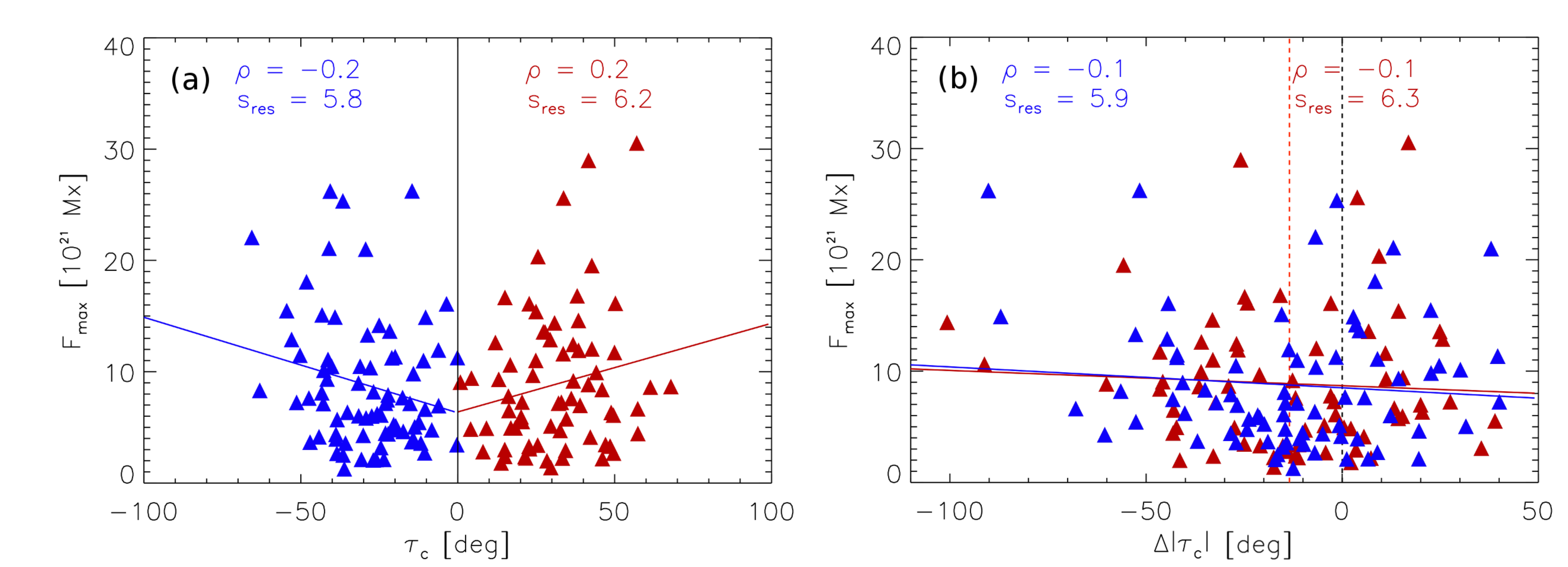}
\caption{(a) Maximum magnetic flux {\vs} $\taucMean$ for 149 ARs during their emergence. Red and blue triangles represent positive and negative values of $\taucMean$, respectively. 
   (b) Maximum magnetic flux {\vs} the change of $|\tauc |$ [$\DelAbstauc$] during the AR emergence phases.  Linear fits for positive (red line) and negative (blue line) twist are shown. For both plots we include the correlation parameter of the linear fit [$\rho$] and the standard deviation of the residuals [$S_{\rm res}$].
   }
 \label{fig:correl(Flux,DTc,Tc)}
\end{center}
\end{figure}

\subsection{Is the Tongue Angle Correlated with the AR Parameters?}
 \label{sec:tauCorrelation}

The absolute value of the mean tongue angle [$|\taucMean |$] has a broad distribution for both signs (\fig{histo-tau}). We ask now if $|\taucMean |$ is related to the global parameters characterizing the ARs, such as the magnetic flux, the size of the magnetic polarities, and the tilt angle.  

\fig{correl(Flux,DTc,Tc)}a shows a weak correlation and a large standard deviation between the AR maximum magnetic flux and $|\taucMean |$. Then, while the mean dependance is the same for both positive and negative helicity, there is, at most, only a weak dependance of $|\taucMean |$ on the AR magnetic flux. Otherwise, we found no correlation between $|\taucMean |$ or $|\taucMax |$ and the distance between the magnetic polarities (the AR size) or their extension at the maximum of the flux emergence (\eg\ for $|\taucMean |$ the correlation parameter is 0.0004 and 0.004 and the standard deviation  
14.7 and 14.4 in function of the AR size and extension, respectively. Similar values are also found for $|\taucMax |$).  Neither $|\taucMean |$ nor $|\taucMax |$ are related to the AR mean tilt nor to the duration of the emergence (\eg\ for $|\taucMean |$: correlation parameter $= 0.07, 0.12$ and standard deviation $= 14.7, 14.6$, for the tilt and duration, respectively). 

Therefore, we conclude for isolated bipolar ARs that $|\taucMean |$ and $|\taucMax |$ are independent of most of the AR characteristics, which implies that the amount and distribution of magnetic twist present in an AR has no measurable effect on other global AR properties.
 
\begin{figure}[ht]
\begin{center}
\includegraphics[width=\textwidth]{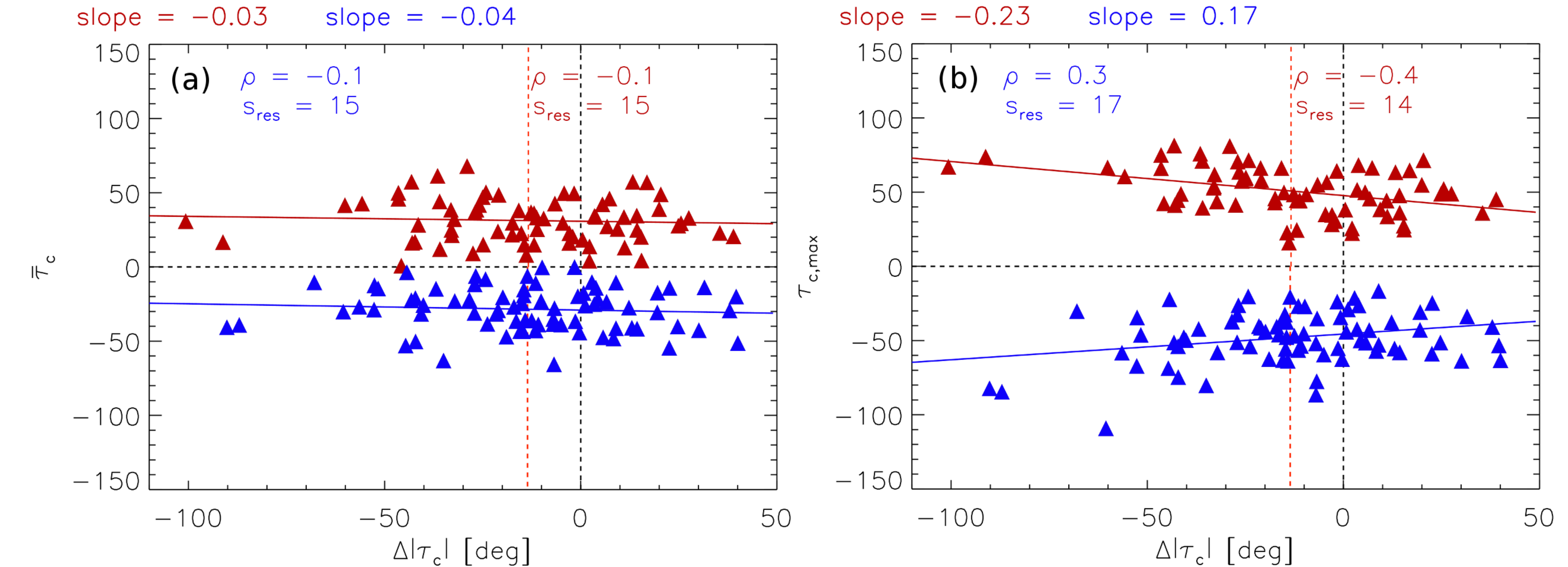}
\caption{(a) Correlation between the mean corrected $\tau$ angle [$\taucMean$] and the change of $|\tauc |$ [$\DelAbstauc$] during the emerging phase. 
(b) Correlation between the signed maximum value of the corrected $\tau$ angle [$\taucMax$] and $\DelAbstauc$. Linear fits for positive (red line) and negative (blue line) $\taucMean$ are shown. For both plots we include the correlation parameter of the linear fit [$\rho$] and the standard deviation of the residuals [$S_{\rm res}$].
}
 \label{fig:correl(Tc,DTc)}
\end{center}
\end{figure}

\begin{figure}[t]	
\begin{center}
\includegraphics[width=.7\textwidth]{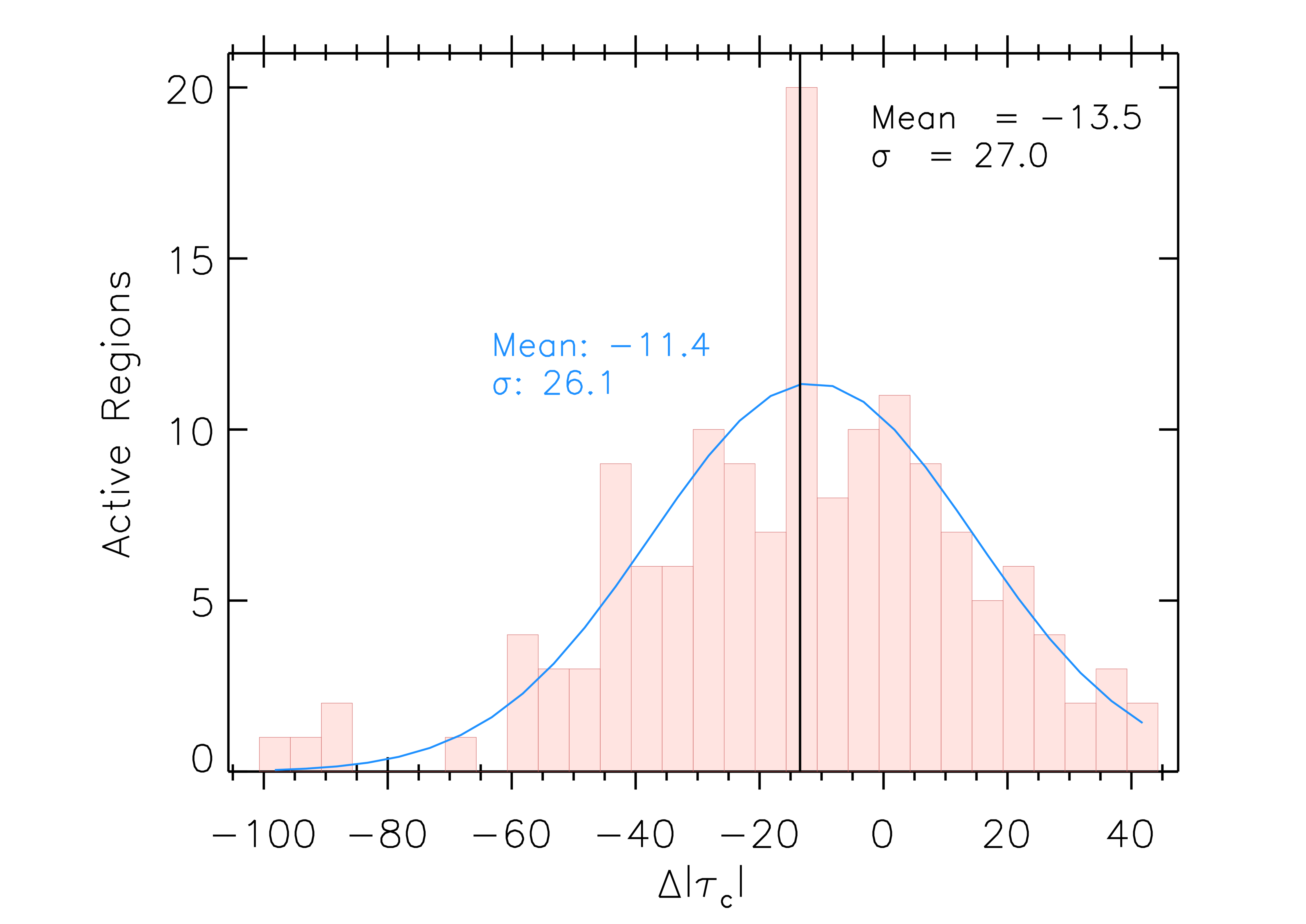}
\caption{Histogram of the change of $|\tauc |$ [$\DelAbstauc$] during the emergence phase of the 149 ARs.  The vertical-black line shows the mean value of $\DelAbstauc$. The distribution is fitted with a Gaussian function using the least-squares method (light-blue line).
}
 \label{fig:histo-dtc}
\end{center}
\end{figure}

\subsection{Evolution of the Tongue Angle During the Emergence}
 \label{sec:Evolution}

During flux emergence $\tauc (t)$ can have very different profiles as a function of time for the ARs studied. Two examples are shown in \fig{example}.  Fluctuations are present in general due to small bipolar emergences that change the PIL locally and quasi-randomly.   These small bipole emergences are at the center of photospheric flux emergence in ARs;  \ie\ they are a consequence of the vertical sharp plasma-pressure gradient at the photospheric level.  These small bipolar emergences are only partially filtered out by the spatial resolution of MDI and by the PIL-fitting method, resulting in $\tauc$ fluctuations, which are occasionally large enough to reverse the sign of $\tauc$.  In this article, we are only analyzing the long-term evolution of $\tauc$ using a linear fit (see \sect{Analyzing}).  

As happens for $|\tauc|$, its slope, or its change during the full emergence period [$\DelAbstauc$ defined in \eq{dtau}] is broadly distributed with no relation with the AR magnetic flux (\fig{correl(Flux,DTc,Tc)}b). 
Furthermore, as for $|\tauc|$, we found no relation, or only a weak one, between $\DelAbstauc$ and other AR parameters, such as $\taucMean$ or $\taucMax$ (\fig{correl(Tc,DTc)}), the phase of the cycle, the AR latitude, the distance between the magnetic polarities or their extension at the maximum of flux emergence, the AR mean tilt, or the duration of the emergence (not shown).

Still, a remarkable result is that $\DelAbstauc$ is almost normally distributed (\fig{histo-dtc}).  Indeed, apart from four outliers with $\DelAbstauc<-80\degree$, the distribution is very symmetric around its median.  The most striking result, however, is a narrow peak located at its mean/median value. The statistical fluctuations around that value are estimated to be $\sqrt{N}\approx \pm 3$ where $N$ is the number of cases per bin. This is corroborated by the fluctuations present on both sides of this peak, while the peak is three times higher than these fluctuations.  Then, a mechanism is present that favors the formation of ARs with this mean $\DelAbstauc$-value, while another mechanism is producing a normally distributed $\DelAbstauc$.  The simplest FR model, the one with a uniform twist (see \sect{FRmodel}), is close to this mean $\DelAbstauc$-value (with values $\approx -5$ to $-7\degree$, see \fig{correl(Dtilt,DTc)}); however, this does not explain why this value should be the most frequent one.

\begin{figure}[t]	
\begin{center}
\includegraphics[width=\textwidth]{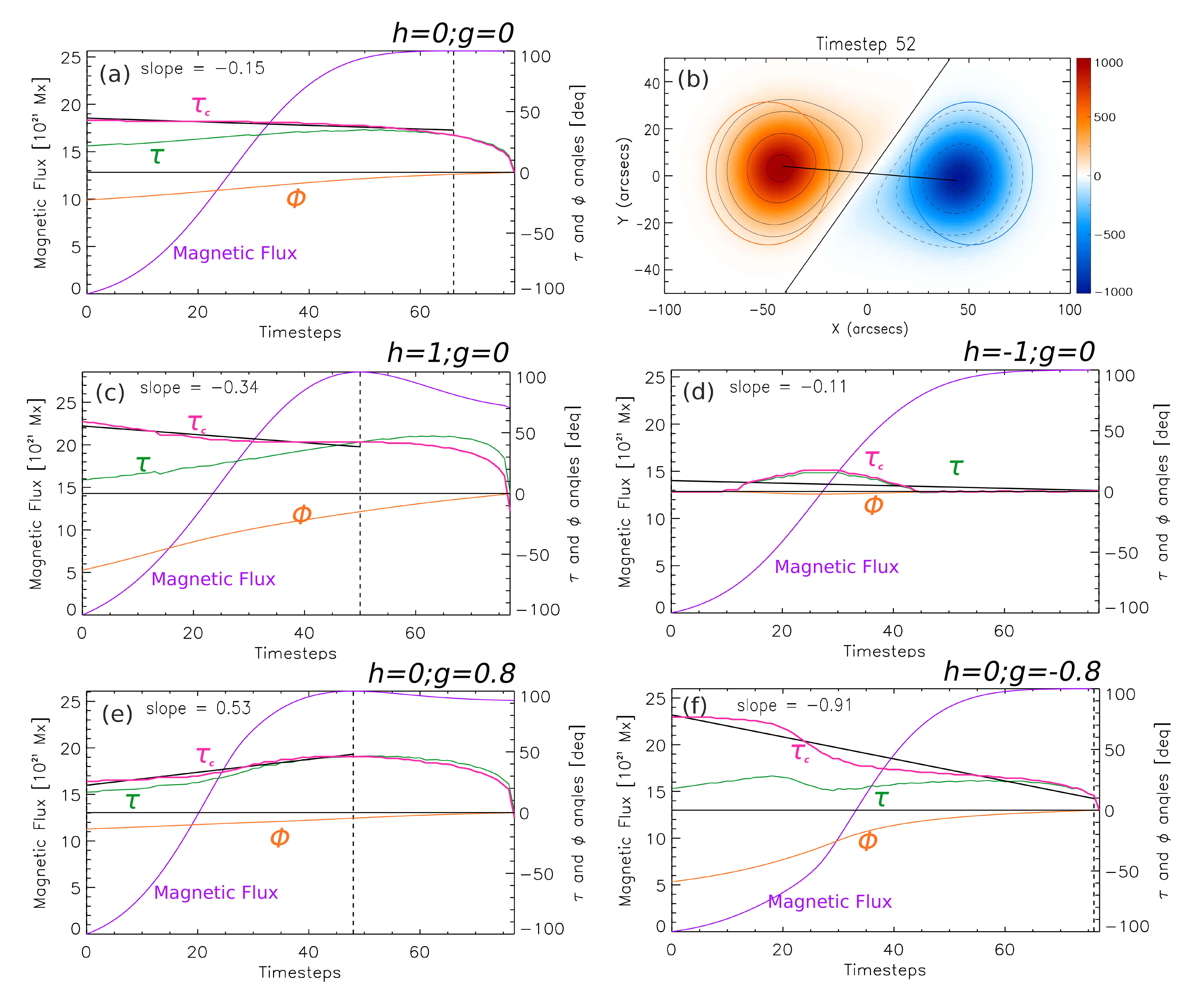}
\caption{ Evolution of the magnetic flux (violet-continuous line), the tongue angle [$\tau$] (green line), its corrected value [$\tauc$] (\eq{tauc}, pink line), and the tilt angle [$\phi$] (orange line), for a toroidal FR model with different $h$- and $g$-parameters (see \append{model}). The vertical-dashed-black line mark the timestep of the maximum magnetic flux (the end of the emergence phase) and the period where we perform the least-squares fit of $\tauc$.
  (a) Basic model with uniform twist (\sect{Brief-Model}; see \href{run:./movie9a.mp4}{\sf{Movie9a}}).
  (c)\,--\,(d) Two extreme cases of the twist profile [$\Nt (\rho)$] (\eq{Nt})(see \href{run:./movie9c.mp4}{\sf{Movie9c}} for $h=1$ and \href{run:./movie9d.mp4}{\sf{Movie9d}} for $h=-1$).  (e)\,--\,(f) Two extreme cases of the vertical dependence of the axial field component (\eq{Bphi})(see \href{run:./movie9e.mp4}{\sf{Movie9e}} for $g=0.8$ and \href{run:./movie9f.mp4}{\sf{Movie9f}} for $g=-0.8$).
  Panel b shows a synthetic magnetogram obtained with a uniformly twisted FR having a positive twist [$\Nto = 0.5$] and an aspect ratio $a/R = 0.4$. The axis that joins the barycenters (straight-thin-black line) is approximately in the $x$-direction and the PIL is indicated with a straight-thick-black line.  The continuous (dashed) lines are isocontours of the positive (negative) magnetic-field component normal to the photosphere, drawn for four different values taken at 10, 25, 50, and 70\% of the axial field strength $B_{\rm max} = 1000$ G. The red- and blue-shaded regions indicate positive and negative polarities, respectively. 
}
 \label{fig:model-results}
\end{center}
\end{figure}

\section{Comparison with a Simple Flux-Rope Model}
 \label{sec:FRmodel}

\subsection{Brief Description of the Basic Model}
 \label{sec:Brief-Model}

\citet{Luoni11} used a simple toroidal FR model to explain the origin of the magnetic tongues in emerging ARs as due to the effect of the azimuthal field-component.  This model was further compared quantitatively to emerging ARs by \citet{Poisson15}. The model was designed to be the simplest possible, while representing the basic characteristics of magnetic tongues.  It consisted of a toroidal FR with uniform twist (both along and across its axis).  The upper half of the torus was set to progressively emerge without distortion.  Therefore, this simple model did not take into account the deformations and reconnections occurring during the emergence.  These reconnection processes, although transforming the internal structure and allowing the emergence (by letting the dense plasma escape), were not expected to change very significantly the global distribution of the magnetic field, \ie\ for the scales of the order or larger than the FR cross-section size.  Therefore, the observed photospheric magnetic field was expected to reflect the global distribution of the FR magnetic field below the photosphere. The emergence of the FR at the photospheric level provide a series of synthetic magnetograms (see an example in \fig{model-results}b), which were analyzed in exactly the same way as observed magnetograms. 

As in \citet{Poisson15}, we link the FR number of turns [$\Nto$] of the field lines around half of the torus azimuthal axis, with the tongue angle [$\tauc$] present in the synthetic magnetograms. The present observations show a broad range of $|\tauc |$, within the interval $[0\degree ,70\degree ]$, corresponding to a broad range of $|\Nto |$, within the interval $[0 ,1.3]$. 

The same parameters as in observations, but derived from the model, are shown in \fig{model-results}a for $\Nto = 0.5$. This case corresponds to the basic model with a
uniformly twisted flux tube, as the one in \citet{Luoni11}.  The correction from $\tau$ to $\tauc$ is relatively important due to the strong magnetic flux present in the tongues.  This induces an apparent rotation of the bipole, hence a change in the tilt $\phi$ as the FR is emerging.  It also implies that $\tauc$ is slightly decreasing with time (even when the twist is uniform).  The $\DelAbstauc$ value for this model, with $\Nto = 0.5$, is $\approx -7\degree$ (see \fig{model-results}a) and is comparable with the mean value obtained in \fig{histo-dtc}.  Changing the value of $\Nto$ in the model \p{ affects mainly} the mean value of $\tauc$; however, by doing so, we can interpret qualitatively only a fraction of the observed ARs. Therefore, our simple model needs to be extended so that a much larger variety of $\tauc$-slopes can be included to represent those shown by the observed ARs (\eg\ compare \fig{model-results}a with panels b and d of \fig{example}).

\subsection{Extension of the Flux Rope Model}
     \label{Extension-Model}

We first test if a non-uniform twist in the FR cross-section, depending on the small radius of the torus [$\rho$] could result in different 
$\tauc (t)$ profiles.  We still consider field lines located on toroidal shapes (\ie~with
$\rho$ independent of $\phi$), then we introduce a dependance of $\Nt$ only with
$\rho$ ($\Nt$ is uniform on each of these tori and changes from torus to torus with
$\rho$).  More precisely, we assume a parabolic shape for $\Nt (\rho)$ controlled
by the unique parameter $h$ (see \eq{Nt} in \append{model-B}).  For $h=0$ the twist is uniform (this corresponds to the case described in \sect{Brief-Model}).  For $h>0$ the twist is more concentrated at the periphery, being twice as large at the FR periphery as at its center for $h=1$.  Conversely, for $h<0$ the twist is maximum at the FR center and it vanishes at the periphery for $h=-1$. The twist is set to zero at larger $\rho$-values as indicated by \eq{Nt}.
  
The mean value of $\tauc$ is affected by the value of $h$ (see \fig{model-results}c\,--\,d), but this can be compensated by changing $\Nto$. In what follows, we keep $\Nto=0.5$ in all figures, unless explicitly stated otherwise.
The global slope (from a linear fit) of $\tauc$ is only weakly affected by $h$ (\eg\ \fig{model-results}c,d).  Indeed, because the twist profile is symmetric around the FR axis (see \eq{Nt}), changing the value of $h$ affects the PIL orientation in a similar way when both the top and the bottom parts of the FR apex 
cross $z=0$.  This twist 
non-uniformity modifies the PIL orientation (and then $\tauc$) similarly at the beginning and at the end of the emergence phase. Therefore, changing the twist profile does not efficiently change the $\tauc(t)$-dependence.   To have a larger variety of $\tauc(t)$-profiles, we would need a twist concentrated either at the top or at the bottom of the FR apex.  However, the very nature of a FR is to have field lines spiraling around its axis satisfying the conservation of the magnetic flux; so, it is difficult to envision a very different azimuthal field component at the top from that at the bottom of a FR. The effect of bending the FR axis is already taken into account by solving $\vec{\nabla} \cdot \vec{B} =0$; this implies a stronger azimuthal field component at the FR bottom (see \eq{Btheta}).

The model discussed in the two previous paragraphs still needs to be extended to understand the variety of observed cases (\eg\ \fig{histo-dtc}). Analyzing different modifications, we find that the most relevant one is to include a dependence of the axial field component on the local vertical direction, more precisely in the direction away from the torus center (\eq{Bphi} in \append{model-B}).  This change introduces a redistribution of the axial field within the flux rope cross-section which satisfies the $\vec{\nabla} \cdot \vec{B} =0$ constraint.  As in the models discussed previously, we do not solve the balance of forces in order to keep them analytically as simple as possible.  We could add a plasma pressure to balance the Lorentz force for $z<0$. For the coronal region, $z>0$, the model would need a numerical relaxation to a force-free field; however this is not required for our purpose since we do not analyze this coronal part in this study.    

The asymmetry of the axial field component is taken into account by simply introducing another parameter [$g$] which is within the interval $[-1,1]$ to avoid a reversal of the axial field in some part of the FR (see \append{model-B}).  The parameter $g$ has an important effect on $\tauc (t)$ as it introduces a strong asymmetry between the top and the bottom parts of the FR cross-section; so, this extension of the model can affect strongly the evolution of the magnetic tongues.
A more negative $g$-value decreases the slope of $\tauc (t)$ (\fig{model-results}f), while a sufficiently large positive value of $g$ can induce a significant positive slope (\fig{model-results}e) as observed for some ARs (\eg\ \fig{example}d, and more generally \fig{histo-dtc}). This trend is general as shown in \sect{Interpretation} and \fig{correl(Dtilt,DTc)}. 

In summary, the new model presented in this section has two new non-dimensional free parameters [$h$ and $g$] which are both typically within the interval $[-1,1]$ in order to produce synthetic magnetograms having the main characteristics of observed emerging ARs. This extension of the basic FR model with uniform twist is the minimum required to interpret all the observations of ARs, as shown in the next section.

\begin{figure}[ht]
\begin{center}
\includegraphics[width=0.7\textwidth]{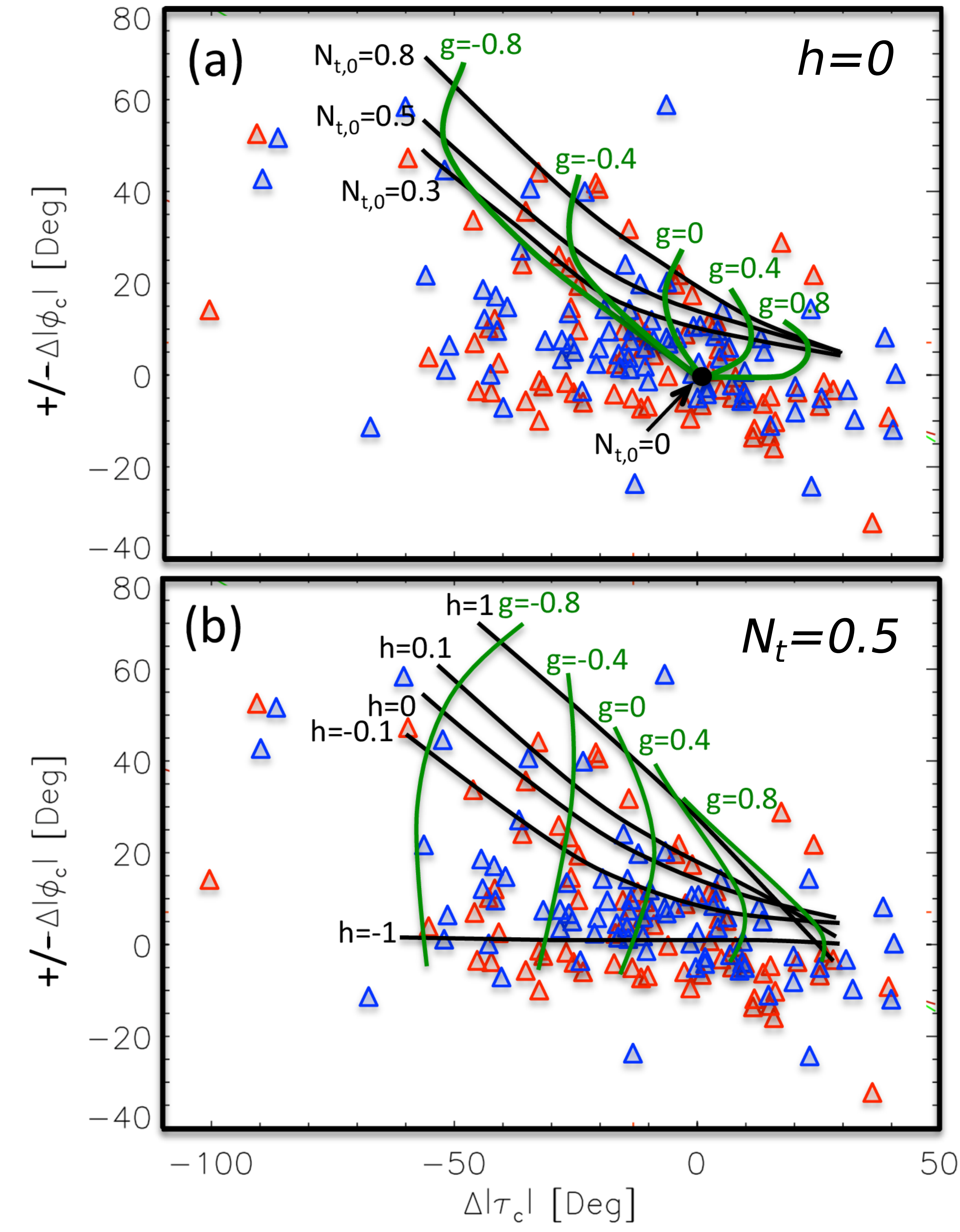}
\caption{  
  Comparison of the observations with the torus model (\append{model}) within the parameter space describing the variations of the tilt angle [$\DelAbsphic$] and of the corrected tongue angle [$\DelAbstauc$] during the emergence phase of ARs.  The ARs with positive (negative) $\tauc$, \ie\ twist, are represented with red (blue) triangles, respectively.  A $-$ ($+$) sign is added in front of $\DelAbsphic$ for positive (negative) twist in order to superpose all the ARs on the same plot.  The results of the model are overlaid with curves having a constant value of the indicated parameter.
  (a) Effect of the number of turns [$\Nto $] in half of the torus (shown in \fig{torus}) and of the parameter $g$ controlling the vertical asymmetry of the axial field (\eq{Bphi}).  All of the models have $\Nt$ independent of $\rho$ ($h=0$).
  (b) Same as (a) but with the effect of a variable twist profile as controlled by the parameters $h$ and $g$ (\eqs{Nt}{Bphi}) for $\Nt =0.5$.
}
 \label{fig:correl(Dtilt,DTc)}
\end{center}
\end{figure}

\subsection{Comparison of the Model to Observations}
\label{sec:Interpretation}

The model described in \append{model} depends on several parameters, such as the torus radius, its thickness, the emergence time-duration, and the axial flux, which can be scaled to any observed AR.  It can also be rotated to any tilt.  The most relevant parameters to have well-defined magnetic tongues are the number of turns around the axis of a half torus [$\Nto$] and the parameters [$h$ and $g$] controlling the azimuthal and axial field components. These latter parameters define the magnitude and the evolution of the relative PIL orientation, measured by $\tauc (t)$, as well as the variation of the tilt angle [$\phic (t)$].

The configurations with $\tauc$ and $-\tauc$, or equivalently $\Nt$ and $-\Nt$, are mirror images one of the other \citep[\eg\ see Figure 1 of ][]{Luoni11}.  They can be compared by plotting $|\taucMean|$ and $\DelAbstauc$.  They also show an opposite variation of the tilt angle due to the retraction of the tongues after they dominated the evolution of the tilt angle during the earlier emergence phase of the ARs.  We define the corrected tilt angle [$\phic$] by subtracting the tilt at the maximum magnetic flux from $\phi$ (\eq{phic}).  Then, as for $\tauc$, we compute $\DelAbsphic $ to better compare ARs with positive and negative magnetic helicity.  \fig{correl(Dtilt,DTc)} shows that ARs with both $\taucMean$ signs are similarly spread in the diagram $\DelAbsphic$  [$\DelAbstauc$].  The same conclusion was reached before when using other AR parameters (\figss{AR(latitude,time)}{correl(Tc,DTc)}).  Then, as expected, emerging ARs have properties that are independent of their magnetic-helicity sign.
   
We first describe the results for the model with uniform twist [$h=g=0$] as shown by the curve $g=0$ in \fig{correl(Dtilt,DTc)}a.  This model can represent only a very small fraction of ARs by varying the axial number of turns [$\Nto$].  The untwisted case, located at the origin of the plot (black dot), has no special relevance for the observations (\ie\ the ARs are not especially clustered around this case).
Introducing a twist profile [$h \ne 0$] allows us to interpret a few other ARs, only a slightly larger fraction along the curve $g=0$ in \fig{correl(Dtilt,DTc)}b. Therefore, combining scans of the parameters $\Nto$ and $h$ with $g=0$ allows us to interpret a small fraction of the emerging ARs; only the ones with a small evolution of $\tauc$ [$\DelAbstauc$ slightly negative in \fig{correl(Dtilt,DTc)}]. 

It is only by varying $g$ within its full allowable range that the variety of results for observed ARs can be reproduced. Varying $g$ in the interval $[-0.8,0.8]$ allows us to reproduce most of the observed range of $\DelAbstauc$ (\fig{correl(Dtilt,DTc)}).
\p{This indicates that there are two extreme cases with a full spectrum in between. In the first case, the axial field strength is stronger at the beginning of the emergence, while in the second case the axial field is stronger at the end.  
This is equivalent to a local twist that is weaker at the top compared to the bottom of the emerging FR apex in the first case, and {\it vice versa} in the second case (see \eq{localTwist} for the local-twist definition).}
Such variety of emerging FRs is intriguing.  We cannot identify any other AR characteristic that can be related to $g$ (\eg\ there is no total-flux dependence of $\DelAbstauc$, as shown in \fig{correl(Flux,DTc,Tc)}, and a possible interaction with other polarities existing before the FR emergence is prohibited by our emerging AR selection, since they emerge in a nearly field-free environment).            
Moreover, as far as we know, this property has never been analyzed in numerical simulations of FR emergence. 

We conclude that the FRs forming isolated bipolar ARs have a broad range of
twist profiles when the observed results are interpreted using the simple model
developed in \append{model}. To explain this, a variable asymmetry between the top and bottom of the FR apex is required. 
In fact, adding a variation of the twist around the FR axis allows the model to reproduce the main characteristics of the observations. In particular, the variation of the magnetic-tongue strengths can be interpreted changing the parameters $h,g$ within the interval [-1,1] (\fig{correl(Dtilt,DTc)}b), but we cannot exclude the presence of a variable axial twist [$\Nto$] (\fig{correl(Dtilt,DTc)}a). In fact, slightly different sets of $\Nto, g, h$ values could describe a particular AR especially because of the interplay between $\Nto$ and $h$.

Extending the range of $\Nto$ to larger values does not significantly extend the range covered by the model in \fig{correl(Dtilt,DTc)} 
(\eg~for $\Nt=1.3$ there is only a small extension to $\DelAbstauc \approx -70$ in the bottom left corner of the covered domain of \fig{correl(Dtilt,DTc)}b). Moreover, such a high value of $\Nto$ creates synthetic magnetograms with large tongues having too much magnetic flux when compared to observed ARs. Extending the ranges of $g$ or $h$ also creates synthetic magnetograms that are not representative of observed ARs (because $B_{\theta}$ or $B_{\rm z}$ change of sign within the flux rope). Then, while the model can describe the distribution of the majority of ARs (62 \% of the ARs are included in \fig{correl(Dtilt,DTc)}b), it cannot describe the ARs at the periphery of this domain. This points again to the limitations of this analytical model.

The broad distribution of $g$ could have its origin in the dynamo process building up FRs or during their transport, especially at the storage phase below the photosphere before emergence. The involved process has an intrinsic random effect shown by the nearly Gaussian distribution of $\DelAbstauc$ (\fig{histo-dtc}), while being inhibited in some cases (narrow central peak of the distribution).  This points to a plausible effect of large convective cells on the studied emerging ARs through a differential diverging/converging effect of convective flows across the FR cross-section. More precisely, the top part of an emerging FR that is more dominantly in a divergent flow pattern than its bottom part would have a lower twist {\it per} unit length at its top part than at its bottom; so, $|\tauc|$ would increase during the emergence. The reverse would happen when the top part is in a more converging flow pattern. This implies that the evolution of $\tauc(t)$ during the emergence phase would depend on the location of the AR emergence within the convective-cell pattern.

\section{Conclusions}
 \label{sec:Conclusions} 

We analyze the emergence phase of bipolar ARs using photospheric line-of-sight magnetograms during Solar Cycle 23 and the beginning of Cycle 24. To have clearer results we select bipolar ARs whose emergence occurred around the central meridian. Our selection included 187 ARs in a temporal range of 13 years.
At least 80\,\% of the studied ARs present clear observable elongations of their magnetic polarities, or tongues, during more than $50$\,\% of their emergence time (when the background flux or any disturbing extra flux was not high enough to affect significantly the magnetic tongues). 
We apply the method described by \citet{Poisson15} to define the mean PIL inclinations of the bipolar ARs. This method defines the tongue angle [$\tauc$] that characterizes the twist of  emerging flux-ropes (FRs) producing ARs.
The method has proven to be efficient in reducing the large amount of data (more than 4000 line-of-sight magnetograms) to a few parameters that characterize the tongue evolution of the ARs.
   
We define the mean and maximum values of $\tauc$, $\taucMean$, and $\taucMax$, during the full emergence period. Both $\taucMean$ and $\taucMax$ have only a weak sign dominance in each solar hemisphere (53 and 58 \% of dominance); so, as in previous studies involving young ARs \citep[\eg\ ][]{Pevtsov14}, the hemispherical rule is weak.  We also study the variation of $\tauc$ during the full emergence, being $\Deltauc$ its total change. We find no relation between the observed tongue characteristics [$\taucMean, \taucMax, \Deltauc$] and the different periods of the solar cycle. The same happens for the amount of magnetic flux, size of the magnetic polarities, latitude, emergence rate, {\it etc.}. Therefore, the helicity in the studied ARs must be generated independently from other FR properties. 

A striking result is that the total change of $\tauc$ [$\Deltauc$] has a Gaussian distribution with an added very narrow peak at the position of its mean value (\fig{histo-dtc}). After comparing the results derived from observations with those of a FR model, we propose that this distribution could be the result of large convective cells having a differential effect between the top and bottom part of the FR cross-section.  This would imply a variation of the local twist {\it per} unit length along the FR axis, as traced by the evolution of $\tauc$ during the FR emergence. 

More generally, we can summarize the magnetic-tongue evolution during an AR emergence with three main parameters: the mean tongue angle [$\taucMean$] and its total change, $\Delta \tau_{\rm c}$, as well as the tilt change [$\Delphic$] during the full emergence phase.
The distribution of the main characteristics of the magnetic tongues [$\taucMean$, $\Delta \tau_{\rm c}$, and $\Delphic $] can be reproduced by the simple analytical model of \append{model}, which extends the previous uniformly twisted model \citep{Luoni11}.  $\taucMean$ is predominantly affected by the amount of twist and its sign corresponds to the magnetic-helicity sign \citep{Luoni11}. $\Deltauc$ and $\Delphic $ are broadly distributed according to the twist magnitude and its distribution within the FR, as follows. 

We search for the minimal model, \ie\ the one with the lowest number of free parameters, that can reproduce the main characteristics of the observations. The main parameters of the model are its axial twist [$\Nto$: the number of turns for a half torus], and two dimensionless parameters [$h$ and $g$] that characterize the twist distribution.
We set the constraint of no reversal for both the azimuthal- and axial-field components within the FR model to avoid the presence of parasitic polarities with unobserved characteristics in the synthetic magnetograms. Within these limits, the scan of the three main parameters of the model [$\Nto$, $h$ and $g$] allows us to describe the observed variety of emerging ARs (\fig{correl(Dtilt,DTc)}). This constrains the type of FRs that form isolated bipolar ARs, as well as it challenges dynamo and transport models to create such a large variety of FRs. However, we cannot define the twist magnitude and its profile independently from the magnetic-tongue evolution since there is an overlap in the observable consequences of twist magnitude and profile (\ie\ changing $\Nto$ and $h$ cover a common portion of the parameter space, see \fig{correl(Dtilt,DTc)}). We can still claim that uniformly twisted flux tubes are not enough to interpret the observations and that profiles with less twist at the periphery ($h<0$) are required to explain a significant part of the present AR observations. 

The simplicity of the analytical model, with no force balance, arising from a special treatment of the divergence-free property of the field and assumed functions for $\Nt(\rho)$ and $\asy(\theta)$, allows us to do a complete scan over the limits of the free parameters with small computational effort. However, it is obvious from the acquired results, as detailed in \sect{Interpretation} and in the conclusions of this article, that a more realistic numerical model with proper full treatment of the equations and relevant forces is necessary to fully explain the properties of bipolar ARs, as the chosen simple analytical model fails to explain 38\,\% of the selected AR sample.

Even more unexpected is the fact that the variety of emerging-AR properties is reproduced only if a large gradient of the axial magnetic field is present or, equivalently, that the FR is differentially twisted at the top and bottom parts of its apex. 
As far as we know, such twist gradient and its variation have never been reported by numerical simulations of emergence and its origin is unknown.  
One possibility is that it is produced during the magnetic-field storage that occurs below the photosphere before the undulatory instability leads to the emergence of the main FR broken into smaller flux tubes.  
A differential effect of flow divergence between the top and bottom parts of the FR apex, due to the flow motion in a large convective cell, could be at the origin of the observed variety of tongue evolutions during AR emergence.  This process needs to be tested using numerical simulations, before any firm conclusions can be achieved.

  \begin{figure}    
   \centerline{\includegraphics[width=\textwidth,clip=]{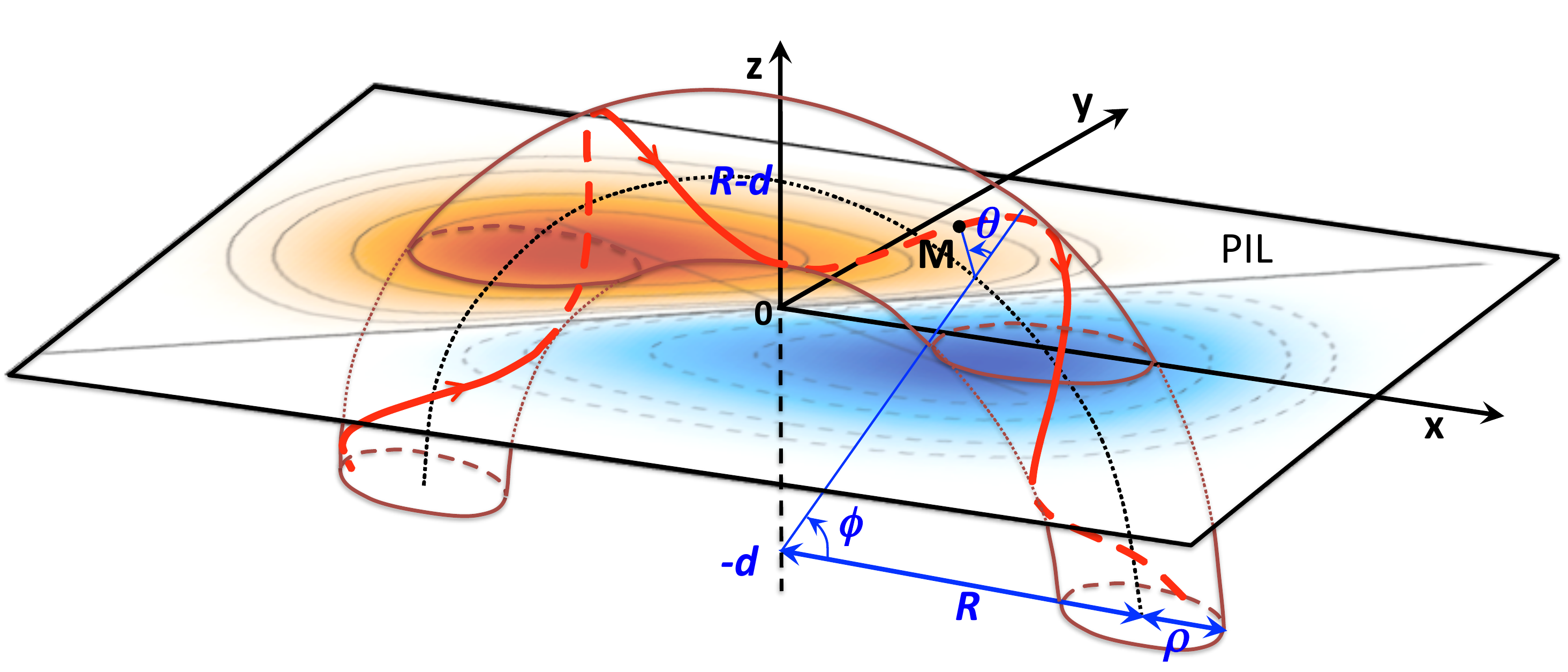}
              }
   \caption{Model of a twisted flux tube having a half-torus shape with a main radius $R$ and a center located at $z=-d$ below the photospheric level ($z=0$).  The twist is positive and uniform ($\Nt =0.5$ turn in half a torus, $h=g=0$ in \eqs{Nt}{Bphi}). The photospheric magnetogram of $B_{\rm z}$ is shown with isocontours and color levels as for observations (\fig{example}) and the superposed brown ellipses (continuous/dashed lines) show the intersection of the FR with $z=-d$ and $z=0$. The torus is outlined for a minor radius $\rho$.
The red line is an example of a magnetic-field line (drawn with an enhanced twist by a factor $\approx 6$ to better outline the FR structure).
 $\phi$ and $\theta$ are the angular coordinates along and around the axis, respectively.}
    \label{fig:torus}
  \end{figure}

%
 \begin{acks}

SOHO is a project of international cooperation between ESA and NASA. MP, MLF, and CHM acknowledge financial support from grants PICT 2012-0973 (ANPCyT), PIP 2012-01-403 (CONICET), and UBACyT 20020130100321 (UBA). MLF and CHM are members of the Carrera del Investigador Cient\'{\i}fico of the Consejo Nacional de Investigaciones Cient\'{\i}ficas y T\'ecnicas (CONICET) of Argentina. MP is a CONICET Fellow.

 \end{acks}


\section*{Disclosure of Potential Conflicts of Interest}

The authors declare that they have no conflicts of interest.

\appendix

\section{Twisted Flux Tube Model}
\label{app:model}  

\subsection{Geometry of the Model}
 \label{app:model-geometry}

  In this section, we describe a simple model to describe the main properties of magnetic tongues in terms of the emergence of an $\Omega$-shaped twisted flux-tube.  The flux-tube shape is half a torus with a main radius $R$ (\fig{torus}). The torus center is located at the height $z=-d$ below the photosphere, which is located at $z=0$. Progressively decreasing $d\geq 0$ simulates a very simplified emergence, \ie\ the FR is emerging without any deformation (in contrast to the results of numerical simulations). Our aim is to illustrate the global, and expected to be robust, implications of the FR twist on the magnetic-tongue evolution; therefore, this very simple model is selected with this purpose.  

Cartesian coordinates, $\{x,y,z \}$, are adopted to describe the model magnetogram located at $z=0$.
However, the natural coordinates for the torus are $\{\rho, \phi, \theta \}$, where $\rho$ is the distance to the torus axis, $\phi$ defines the location along the flux tube axis, and $\theta$ corresponds to the rotation angle around the flux-tube axis (\fig{torus}).  These are the toroidal curved cylindrical coordinates that we simply call torus coordinates below. 
In the axis plane, the unit vector normal to the axis is denoted as $\uR$ and the unit vector along the axis (toroidal direction) is denoted as $\up$. 
Finally, the unit vector around the FR (poloidal direction) is noted $\ut$ and 
the unit vector along the $\rho$-direction, so both normal to the axis and to $\ut$, is noted $\urho$.
Their $\{x,y,z \}$ components are:
  \BA 
  \uR   & = & \{ \cos \phi , 0 , \sin \phi \}  \nonumber \\
  \up   & = & \{-\sin \phi , 0 , \cos \phi \}  \nonumber \\
  \urho & = & \{ \cos \phi \cos \theta , \sin \theta , \sin \phi \cos \theta\} 
                   \label{eq:unit-vect} \\
  \ut   & = & \{-\cos \phi \sin \theta , \cos \theta ,-\sin \phi \sin \theta\} \nonumber 
  \EA

The \torus\ coordinates need to be transformed to Cartesian ones to obtain synthetic photospheric magnetograms.
A point $M$ in the torus is located at $\vec{OM} = R ~\uR + \rho ~\urho -d ~\uz $
(with $O$ being at the torus center). Then, in Cartesian coordinates,   
   \BA  
   \vec{OM} & = & \{x,y,z \} \nonumber \\
            & = &\{ (R+\rho \cos \theta ) \cos \phi    ,\,
                       \rho \sin \theta                ,\,
                    (R+\rho \cos \theta ) \sin \phi -d  \}  \label{eq:OM} \,,
   \EA 
which provides the transformation of \torus\ coordinates to Cartesian ones and the reverse transformation after some algebraic computations.


\subsection{Definition of the Magnetic Field}
 \label{app:model-B}

The magnetic field [$\vec{B}$] satisfies $\vec{\nabla} \cdot \vec{B} =0$. In \torus\ coordinates, is written as

   \BA  
   \vec{\nabla} \cdot \vec{B} &=& \left( \pder{\rho ~(R+\rho \cos \theta) ~\Brho}{\rho} 
           + \pder{(R+\rho \cos \theta) ~\Bt}{\theta} 
           + \rho \pder{ \Bp}{\phi} \right)  \nonumber \\ 
       &&  \Big/\Big( \rho ~\big( R+\rho \cos \theta \big) \Big) = 0  \,. \label{eq:divB}
   \EA
In order to derive a simple analytical model we do not specify a force balance.
We rather assume that field lines are located on torus shapes with $\rho$ independent of $\phi$.
This implies $\Brho=0$.  Next, we cancel separately the two remaining terms in $\vec{\nabla} \cdot \vec{B} =0$, so $\Bt ~\ut$ and $\Bp ~\up$ are both divergence-free. This separate cancellation is only a special case selected to minimize the complexity of the following analytical derivation that poses certain limitations and implications discussed in \sects{Interpretation}{Conclusions}.  This implies $\Bt = f_{\theta} (\rho,\phi) / (R+\rho \cos \theta)$ and $\Bp = f_{\phi} (\rho,\theta)$ where $f_{\theta}$ and $f_{\phi}$ are two general functions.

The dependence on $f_{\theta}$ in $\phi$ would introduce an asymmetry between the two legs, but we do not consider this asymmetry here, so $f_{\theta} (\rho)$. 
If $f_{\phi}$ is only a function of $\rho$, the integration of field lines implies $ B_{\theta} = 2 \, \rho \, \Nt(\rho)  \, B_{\phi} / (R+\rho \cos \theta)$ where $\Nt(\rho)$ is the number of turns in half the torus of small radius $\rho$.  Following previous studies \citep[see, \eg, ][]{Emonet98} the axial field is typically defined as $B_{\phi} = B_{0} \exp (-(\rho/a)^2)$, where $a$ defines the FR thickness and $B_{0}$ the field strength on the axis, then the azimuthal field is defined as  
   \BE  \label{eq:Btheta}
   \Bt (\rho,\theta) = 2 \, \rho ~\Nt(\rho) \, B_{0} ~e^{-(\rho/a)^2} 
                       ~/ (R+\rho \cos \theta)  \,.
   \EE

We explore the effect of a non-uniform twist profile by defining
   \BE  \label{eq:Nt}
    \Nt(\rho) = \Nto ~ {\rm max} (1+h~(\rho/a)^2,0)  \,, 
   \EE
where $h$ is a parameter. For $h=0$, the FR is uniformly twisted. A value of $h>0$ implies a twist more concentrated at the edge; as an example, if $h=1$ the twist at $\rho=a$ duplicates its value at the center. Similarly, $h<0$ implies a twist decreasing from the axis to the FR border. 
Taking the maximum of the parenthesis in \eq{Nt} avoids the change of sign of $\Nt(\rho)$ within the FR, which would produce strong magnetic tongues with opposite sign to that in the core, a case typically not observed. We further set $h\geq -1$ so that the strong field, within $\rho<a$, is not affected.

For the axial component [$\Bp$] \eq{divB} allows a general form with a $\theta$-dependence (with $f_{\phi} (\rho,\theta)$). Since $\Bt$ and $\Bp$ are independent (no force balance is solved), we exploit this possibility by introducing the simplest $\theta$-dependence needed for our purpose:       \BA  
   \Bp (\rho,\theta) &=& B_{0} ~ \asy (\theta) ~e^{-(\rho/a)^2}      
                         \,, \nonumber \\
   \mbox{\rm with}~~ \asy (\theta) &=& (1+g \cos \theta) 
                         \,, \label{eq:Bphi}
   \EA
where $g$ is a parameter controlling the non-axisymmetric level of $\Bp$. The term in $\cos \theta$ introduces a linear spatial variation of $\Bp$ in the direction $\uR$ (orthogonal to the FR axis).  It has also the property of preserving the total flux of $\Bp$; therefore, changing the value of $g$ only redistributes $\Bp$ within the FR.

A field line is defined by the curve tangent to $\vec{B}$, so where $\rmd \vec{OM}$ is parallel to $\vec{B}$, with:
   \BE  \label{eq:dOM}
   \rmd \vec{OM} = \rmd \rho \,\urho + \rho \,\rmd \theta \,\ut 
                 + (R+\rho \cos \theta ) \,\rmd \phi \,\up  \,.
   \EE 
Then, the field line equation can be written:
   \BE  \label{eq:localTwist}
   \frac{\rmd \theta}{\rmd \phi} 
   = \frac{R+\rho \cos \theta}{\rho} \frac{\Bt}{\Bp}
   = \frac{2 \Nt(\rho)}{1+g \cos \theta}         \,.
   \EE 
This implies that the local twist [$\rmd \theta/\rmd \phi$] is lower in the top part ($\theta \approx 0$) than in the bottom one ($\theta \approx \pi$) of the FR for $g>0$ and the reverse for $g<0$.


\subsection{Synthetic Magnetogram}
 \label{app:model-magnetogram}

The model magnetogram is defined at the height $z=0$ as:
   \BE  \label{eq:Bz}
     B_z(x,y) = B_{0}\, (x ~\asy - 2 ~y ~\Nt ~d ~/u) \,\, \mathrm{e}^{-(\rho/a)^2} \, /u   \,,
   \EE
with $u=\sqrt{x^2+d^2}$.   As $d$ decreases from $d \approx R+a$ to $d \approx 0$,
$B_z(x,y)$ describes the evolution of a theoretical magnetogram where a 
twisted $\Omega$-shaped flux tube is emerging without deformation. 

From \eq{Bz} the PIL of $B_z$ is
   \BE  \label{eq:PIL}
   x = 2  ~y ~\Nt ~d ~/(\asy \sqrt{x^2+d^2})   \,.
   \EE
For the central part of the magnetogram, so for $|x|<<d$, \eq{PIL} is approximately
   \BE  \label{eq:PILapprox}
   x \approx 2  ~y ~\Nto ~(1+h~\rho^2/a^2) / (1+g \cos \theta)   \,,
   \EE
where $\rho$ and $\theta$ can be expressed in function of $(x,y,d)$ (and $z=0$).
The PIL equation further simplifies to 
   \BE  \label{eq:PILapprox0}
   x \approx 2 ~y ~\Nto  \qquad \mbox{or} \qquad  \tan \tauc = 2 ~\Nto \,,
   \EE
for a uniform twist [$h=g=0$]. This limit corresponds to the model used in \inlinecite{Luoni11}.
For a uniform twist, the PIL is straight and $\tauc$ is constant during the emergence (\eq{PIL} implies $x \propto y$ for $|x| \ll d$, so around the central part of the bipole with a spatial extension decreasing with $d$, then during the emergence).   
For a non-uniform twist [$h \ne 0$] and/or an asymmetric twist [$g \ne 0$] across the FR, the PIL is curved according to the twist profile as described by \eq{PILapprox}.  It is also function of $d$, \eq{PIL}, so the PIL evolves during the emergence.  The model results are described in \sect{FRmodel}.



\bibliographystyle{spr-mp-sola} 
\bibliography{paper_tongues} 
\IfFileExists{\jobname.bbl}{}
{\typeout{}
\typeout{****************************************************}
\typeout{****************************************************}
\typeout{** Please run "bibtex \jobname" to obtain}
\typeout{** the bibliography and then re-run LaTeX}
\typeout{** twice to fix the references!}
\typeout{****************************************************}
\typeout{****************************************************}
\typeout{}
}

\end{article} 
\end{document}